\documentclass[12pt]{article}
\usepackage[english]{babel}
\usepackage[cp1251]{inputenc}
\usepackage[a4paper, left=25mm, right=15mm, top=20mm, bottom=20mm]{geometry}
\usepackage{graphicx}
\usepackage{amssymb}
\usepackage{xcolor}
\usepackage{hyperref}
\hypersetup{pdfstartview=FitH, linkcolor=linkcolor,citecolor=citecolor,colorlinks=true} 
\usepackage{amsmath}
\usepackage{setspace}
\usepackage{bm}
\usepackage{cite}
\usepackage{url}
\usepackage[hang,flushmargin]{footmisc} 
\newcommand{\be}{\begin{equation}}
\newcommand{\ee}{\end{equation}}
\renewcommand{\baselinestretch}{1.3}
\date{}
\begin{document}
\definecolor{linkcolor}{HTML}{008000}
\definecolor{citecolor}{HTML}{4682B4}
\large
\title{\begin{spacing}{2}\begin{center}\textcolor[HTML]{696969}{\large REVIEW OF MODERN PHYSICS}\end{center}\end{spacing}
\begin{spacing}{1}\LARGE \bf Theory for the beam splitter in quantum optics: quantum entanglement of photons and their statistics, HOM effect.\end{spacing}}
\normalsize
\author{D.N. Makarov$^{*}$\\
Northern (Arctic) Federal University, Arkhangelsk, 163002, Russia\\
E-mail:  $^{*}$makarovd0608@yandex.ru }
\maketitle
\begin{abstract}
\begin{spacing}{1}
The theory of the beam splitter (BS) in quantum optics is well developed and based on fairly simple mathematical and physical foundations. This theory has been developed for any type of BS and is based on the constancy of the reflection coefficients $R$ (or the transmission coefficient, where $R+T=1$) and the phase shift $\phi$. It has recently been shown that the constancy of these coefficients cannot always be satisfied for a waveguide BS, where $R$ and $\phi$ depend in a special way on photon frequencies. Based on this, this review systematizes the concept of BS in quantum optics into ``Conventional'' and frequency-dependent BS, and also presents the theory of such BS. It is shown that the quantum entanglement, photon statistics at the output ports, and the  Hong-Ou-Mandel (HOM) effect for such BS can be very different. Taking into account the fact that the waveguide BS is currently acquiring an important role in quantum technologies due to the possibility of its miniaturization, this review will be useful not only for theoreticians, but also for experimenters.\\
\end{spacing}

\end{abstract}

\section{Introduction}
The beam splitter (BS) is one of the main devices not only in classical optics, but also in quantum optics. A beam splitter is an optical device that splits a beam of light into a transmitted and a reflected beam. This is the most important device of many optical and measuring systems. For example, such systems can be interferometers: Michelson-Morley, Mach-Zehnder and Hong-Ou-Mandel \cite{Mandel_1995,Scully_1997,Loudon_2000,HOM_1987}. Despite its simple purpose - to separate the incident beam, the beam splitter in quantum optics has a much broader meaning \cite{Mandel_1995,Scully_1997}. In quantum optics, two modes of the electromagnetic field are usually considered (two input and output ports), because even if 1 input port remains unused, it should be considered as an input for vacuum fluctuations \cite{Agarwal_2013}. The main value in quantum optics is the quantum states of the electromagnetic field at the output ports of the beam splitter. Depending on the reflection coefficient $R$ (similar to the transmission coefficient $T$, where $R+T=1$) and the input states of the electromagnetic field, the quantum states of interest can be obtained at the output ports of the BS. This can be used in many applications of quantum technologies. For example, BS is used in linear optical quantum computing Indeed, Knill et al. showed in 2001 that it is possible to create a universal quantum computer using only BS, phase shifts, photodetectors and single photon sources (KLM protocol)\cite{Knill_2001}. Also, using BS, you can create quantum entanglement between the input modes of electromagnetic fields \cite{HOM_1987,Pan_2012,Sangouard_2011}, simulate quantum transport \cite{Nicholas_2017} and determine the degree of identity of photons \cite{HOM_1987,Tambasco_2018}, etc. BS is an integral part of quantum metrology \cite{Pezze_2018} and quantum information \cite{Weedbrook_2012}, including two- and multi-photon interference \cite{HOM_1987,Ou_2007}.

\begin{figure}[!h]
\center{\includegraphics[angle=0, width=0.9\textwidth, keepaspectratio]{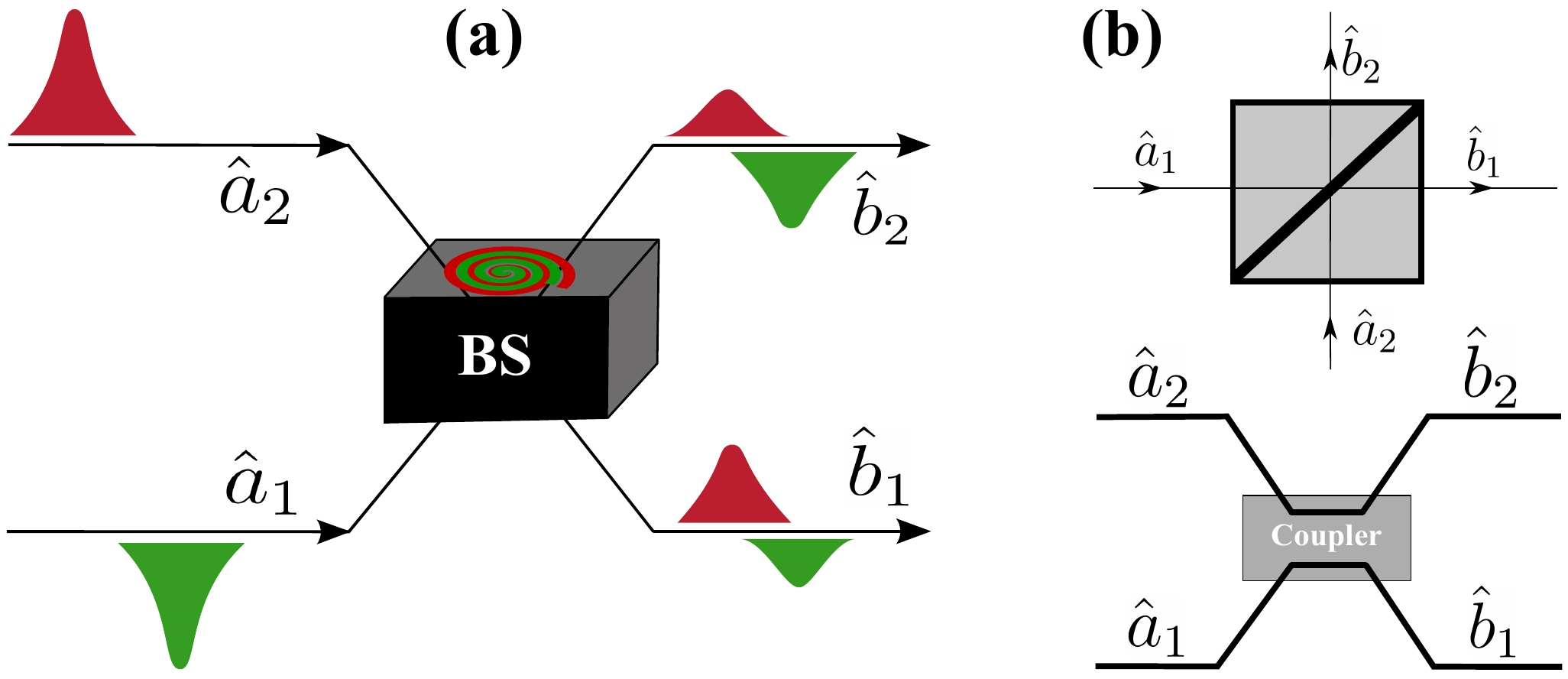}} 
\caption[fig_1]{In Fig.1(a) shows a BS scheme with two input ports and two output ports, where annihilation operators 1 and 2 modes on the input ports represent $\hat{a}_1$ and $\hat{a}_2$ respectively, and on the BS output ports represent $\hat{b}_1$ and $\hat{b}_2$ respectively. The BS is represented as a ``black box'' in which ``mixing'' of electromagnetic field input modes takes place. Fig.1(b) shows the BS with free space optics, i.e., cubic BS (top) and fibre optics, i.e., waveguide BS (bottom).}
\label{fig_1}
\end{figure}

In this review, we will consider two-port beam splitters, since they are the most important and frequently used in quantum technologies. It is well known that such beam splitters can be of various types and properties. By types, BS are divided into ways of their manufacture, for example, a cubic BS or a waveguide BS, see Fig. \ref{fig_1}. The BS can be a glass plate with a silver or dielectric coating, a glass cube with a coating in a diagonal plane, two glass plates with parallel planes, between which a coating is located, or a coating deposited on films. A waveguide BS is two waveguides brought close enough together so that the electromagnetic fields overlap; in this case it is a directional coupler (eg \cite{Bromberg_2009, Politi_2008}). Waveguide BS have an advantage over many types of beam splitters because they are much smaller than them, and also have many other advantages \cite {Tan_2019, Pan_2012, Nicholas_2017}. Subdivision by properties usually means the use of polarized or non-polarized BS or other unique properties. In quantum optics, the BS can generally be represented as (independent of the type of BS) Fig.\ref{fig_1}(a), and the schematic designations depend on their types, e.g. Fig.\ref{fig_1}(b), see \cite{Makarov_SR1_2021}. 
Without going into the technical details and properties of these beam splitters (there are quite a few reviews on this topic, see eg. \cite{Pan_2012}), we can say that in quantum optics there are two main parameters in BS, this is its reflection coefficient $R$ (or transmission coefficient $T$) and phase shift $\phi$. Usually, to study the properties of electromagnetic waves (hereinafter referred to as photons) at the output ports of the BS or in technical devices where the BS is a component, in quantum optics it is considered that $R$ and $\phi$ are constant values, see for example, \cite{Zeilinger_1981,Campos_1989, Kim_2002, Pan_2012, Sangouard_2011, Pezze_2018, Makarov_PRE_2020}.
This means that for finding quantum states of photons at the BS output ports these quantities do not change, i.e. their definite values are always given. For example, in the Hong-Oy-Mandel (HOM) effect these values $R=T=1/2$ and the effect is independent of phase $\phi$. Of course, these values $R$ and $\phi$ depend in general on the wavelength (or frequency) of the incident field (regardless of the BS type), i.e. $R=R(\lambda)$, $\phi=\phi(\lambda)$, then this dependence should have no effect on photon states at the BS output ports if we fix their values. This has always seemed an obvious fact and has not been studied before. Anyway, in general case $R$ and $\phi$ are not constant quantities and the dependence of $R$ and $\phi$ on frequencies is such that it can affect the photon states at the BS output ports. If $R$ and $\phi$ depend on frequencies in this way, then the quantum states at the BS output ports differ with respect to the case of quantum states at constant $R$ and $\phi$. It follows that quantum entanglement at such BS will be different from the case with constants $R$ and $\phi$. As it turns out, such BS is a waveguide beam splitter (the term ``Fiber-optic splitter'' is often used) which differs from many other types of BS by this property.

Recently \cite{Makarov_SR1_2021, Makarov_SR2_2021, Makarov_Enropy_2022} the theory of a frequency dependent waveguide BS has been presented. In these papers, it was shown that if the BS is represented as a coupled waveguide, then the coefficients $ R $ and $ T $ depend on the frequencies of photons entering both ports of the BS. Taking into account the frequency dependence of the coefficients $ R $ and $ T $, many known theories can be modified, for example, the HOM interference theory \cite{Makarov_OL_2020, Makarov_SR_2020} or quantum entanglement of photons based on a beam splitter \cite{Makarov_SR2_2021, Makarov_Enropy_2022}. It should be added that such a frequency dependence of the coefficients $ R $ and $ T $ is inherent only in a waveguide beam splitter.

Thus, there is a need for a review on this topic, where BSs in quantum optics will be systematized into frequency-dependent and non-frequency-dependent ``conventional'' and based on such systematization, the quantum entanglement of photons at the output ports of the beam splitter is considered.
In this review, such a systematization is carried out and not only the quantum entanglement of photons on such beam splitters is considered, but also the statistical properties of photons at the output ports and the HOM effect.

\section{Beam splitter in quantum optics}
Since the BS produces separation of incoming beams, the quantum state of photons at the BS output ports is $|\Psi_{out}\rangle=e^{i{\hat H}t_{BS}}|\Psi_{in}\rangle$, where ${\hat H}$ is the Hamiltonian of the quantized electromagnetic field interacting with matter, $t_{BS}$ is interaction time, $|\Psi_{in}\rangle$ is the initial state of electromagnetic field. It should be added that ${\hat H}$ can be quite complex depending on the type of BS. In general, $|\Psi_{in}\rangle$ can be represented as \cite{Glauber_1965,Ou_2007}
\begin{eqnarray}
|\Psi_{in}\rangle=\sum_{s_1,s_2}\frac{C_{s_1,s_2}}{\sqrt{s_1! s_2!}}{{\hat a_1}^{\dagger}}{}^{s_1} {{\hat a_2}^{\dagger}}{}^{s_2}|0\rangle_1 |0\rangle_2 ,
\label{1}
\end{eqnarray}
where the 1st and 2nd mode creation operators respectively represent ${\hat a_1}^{\dagger}$ and ${\hat a_2}^{\dagger}$, $s_1$ and $s_2$ are the quantum numbers of the 1st and 2nd modes, respectively (or the number of photons in the modes), $C_{s_1,s_2}$ are the expansion coefficients defining the initial state, $|0\rangle_1 |0\rangle_2$ the vacuum states for modes 1 and 2 respectively (for convenience we will write $|0\rangle_1 |0\rangle_2\to |0\rangle$). It should be added that if the initial states are in the Fock state, then the coefficient $C_{s_1,s_2}=1$. In this case, it is easy to show (up to an insignificant phase) that \cite{Glauber_1965, Mandel_1995,Ou_2007}
\begin{eqnarray}
|\Psi_{out}\rangle=\sum_{s_1,s_2}\frac{C_{s_1,s_2}}{\sqrt{s_1! s_2!}}{{\hat b_1}^{\dagger}}{}^{s_1} {{\hat b_2}^{\dagger}}{}^{s_2}|0\rangle,~~{\hat b_k}^{\dagger}=e^{i{\hat H}t_{BS}}{\hat a_k}^{\dagger}e^{-i{\hat H}t_{BS}}~ (k=1,2), 
\label{2}
\end{eqnarray}
where ${\hat b_1}^{\dagger}, {\hat b_2}^{\dagger}$ are the creation operators at the output ports of the BS for modes 1 and 2, respectively. It turns out that these operators can be found in general terms without even considering a particular type of BS, just on the basis of general physical considerations.

It is well known that lossless two-mode BS (with two input and output ports) in quantum optics is described by the unitary matrix $U_{BS}$, which has the form \cite{Agarwal_2013,Zeilinger_1981,Campos_1989,Luis_1995}
\begin{eqnarray}
\begin{pmatrix}
  \hat{b}_1\\\
  \hat{b}_2
\end{pmatrix}=
U_{BS}
\begin{pmatrix}
  \hat{a}_1\\\
  \hat{a}_2
\end{pmatrix},
~~~
U_{BS}=\begin{pmatrix}
\sqrt{T}& e^{i\phi}\sqrt{R}\\\
  - e^{-i\phi}\sqrt{R}& \sqrt{T}
\end{pmatrix} ;
\label{3}
\end{eqnarray}
where the 1st and 2nd mode annihilation operators respectively represent $\hat{a}_1$ and $\hat{a}_2$, and at the BS output ports $\hat{b}_1$ and $\hat{b}_2$; $T$ and $R$ are the transmittance and reflectance, respectively ($R+T=1$), and $\phi$ is the phase shift.  
It should be added that in the literature one often sees the use of the BS matrix $U_{BS}$ in various representations. The most commonly used representation is when the phase shift $\phi=\pm \pi/2$, the second representation often encountered is when $\phi=0$, in these cases the $U_{BS}$ matrix is respectively
\begin{eqnarray}
U_{BS}=\begin{pmatrix}
\sqrt{T}& \pm i\sqrt{R}\\\
  \pm i \sqrt{R}& \sqrt{T}
\end{pmatrix},
~~~
U_{BS}=\begin{pmatrix}
\sqrt{T}& \sqrt{R}\\\
  - \sqrt{R}& \sqrt{T}
\end{pmatrix} .
\label{4}
\end{eqnarray}
In fact, both representations can only be used when the result is independent of the phase shift $\phi$. As will be shown below, many characteristics studied in quantum optics, such as the quantum entanglement of photons at the BS output ports, do not really depend on the phase. In spite of this we will use the general Eq. (\ref{4}).

It should be added that in reality photons are not monochromatic and the frequency distribution \cite{Mandel_1995,Agarwal_2013} must be taken into account. In this case the initial wave function of photons will be 
\begin{eqnarray}
|\Psi_{in}\rangle=\sum_{s_1,s_2}\frac{C_{s_1,s_2}}{\sqrt{s_1! s_2!}}\int \phi(\omega_1, \omega_2){{\hat a_1}^{\dagger}}{}^{s_1} {{\hat a_2}^{\dagger}}{}^{s_2}|0\rangle d\omega_1 d\omega_2 ,
\label{5}
\end{eqnarray}
where $ \phi(\omega_1, \omega_2) $ is the joint spectral amplitude (JSA) of the two-mode wave function. Given the frequency distribution, the output is ($ \int | \phi (\omega_1, \omega_2) |^2 d \omega_1 d \omega_2 = 1 $)
\begin{eqnarray}
|\Psi_{out}\rangle=\sum_{s_1,s_2}\frac{C_{s_1,s_2}}{\sqrt{s_1! s_2!}}\int \phi(\omega_1, \omega_2){{\hat b_1}^{\dagger}}{}^{s_1} {{\hat b_2}^{\dagger}}{}^{s_2}|0\rangle d\omega_1 d\omega_2 .
\label{6}
\end{eqnarray}
\subsection{Basic expressions for beam splitters of any type}
Consider the beam splitter in general terms without defining its type. Let us set the problem to find the matrix $U_{BS}$ on the basis of general physical considerations. This problem in quantum optics has been solved quite a long time ago, see eg \cite{Zeilinger_1981,Campos_1989,Luis_1995}. The result has essentially always been the same - this BS matrix has been obtained in the Eq.(\ref{3}). Let us present such a conclusion here in the simplest version.
For this purpose we will disregard the effects of polarisation, misalignment and imperfect beam collimation, as these can be incorporated into BS theory quite easily.  The input (${\hat a_1}, {\hat a_2}$) and output (${\hat b_1}, {\hat b_2}$) annihilation operators, at some chosen frequency for linear BS can be represented as
\begin{eqnarray}
\begin{pmatrix}
  \hat{b}_1\\\
  \hat{b}_2
\end{pmatrix}=
\begin{pmatrix}
U_{11}& U_{12}\\\
U_{21}& U_{22}
\end{pmatrix}
\begin{pmatrix}
  \hat{a}_1\\\
  \hat{a}_2
\end{pmatrix}.
\label{7}
\end{eqnarray}
It should be added that the representation as Eq.(\ref{7}) is a direct consequence of the conservation of particles (photons) before and after their passage through BS. If photons before hitting the BS were $s_1+s_2$ and after passing through the BS became $n+m$, then for the number of photons to be conserved $s_1+s_2=n+m$ requires that the transformation matrix be in the form Eq. (\ref{7}) \cite{Glauber_1965,Campos_1989}. The conversion matrix $U_{BS}$ contains the elements $U_{i,j}$ ($i,j=1,2$), which are generally complex and must be represented as $U_{i,j}=|U_{i,j}|e^{i\phi_{i,j}}$. The well-known bosonic commutation relations at the output of the BS must also hold, i.e. $[\hat{b}_i,{{\hat b_j}^{\dagger}}]=\delta_{i,j}$. This commutation relation leads to the relations
\begin{eqnarray}
|U_{1,1}|^2+|U_{1,2}|^2=1,~~|U_{2,1}|^2+|U_{2,2}|^2=1,~~U_{1,1}U^{*}_{2,1}+U_{1,2}U^{*}_{2,2}=0.
\label{8}
\end{eqnarray}
The last relation in Eq. (\ref{6}) can be simplified by reducing to 2 equations, we get
\begin{eqnarray}
|U_{1,1}| |U_{2,1}|=|U_{1,2}||U_{2,2}|, ~~ \phi_{1,1}-\phi_{1,2}=\phi_{2,1}-\phi_{2,2}\pm \pi.
\label{9}
\end{eqnarray}
As a result, from the first two ratios in Eq.(\ref{8}) and from the first relation in Eq.(\ref{9}) it can be deduced that $|U_{2,1}|^2=|U_{1,2}|^2=R=\sin^2 \theta,~~|U_{1,1}|^2=|U_{2,2}|^2=T=\cos^2 \theta$, where $0 \leq \theta \leq \pi/2$. In this case, one can see that the relation $R+T=1$. From the point of view of physics, these coefficients $R$ and $T$ are the coefficients of reflection and transmission, respectively. The four-phase dependence can be reduced to a two-phase dependence by making phase substitutions in the form: $\phi_{\tau}=1/2(\phi_{11}-\phi_{22}); \phi_{\rho}=1/2(\phi_{12}-\phi_{21}\mp \pi); \phi_{0} =1/2(\phi_{11}+\phi_{22})$. In this case, the BS matrix Eq. (\ref{7}) will be
\begin{eqnarray}
U_{BS}=e^{i\phi_0}\begin{pmatrix}
\cos \theta e^{i\phi_{\tau}} & \sin \theta e^{i\phi_{\rho}}\\\
-\sin \theta e^{-i\phi_{\rho}}& \cos \theta e^{-i\phi_{\tau}}.
\end{pmatrix} 
\label{10}
\end{eqnarray}
In Eq. (\ref{10}) has only two phases left, since $\phi_0$ is the phase multiplier, which is known to be negligible. We can see that by using the BS matrix in the form of Eq. (\ref{10}), the operators $\hat{b}_1, \hat{b}_2$ can be further simplified, we obtain:  $\hat{b}_1=e^{i\phi_{\tau}}(\cos \theta \hat{a}_1+\sin \theta e^{i(\phi_{\rho}-\phi_{\tau})}\hat{a}_2)$;~$\hat{b}_2=e^{-i\phi_{\tau}}(-\sin \theta e^{-i(\phi_{\rho}-\phi_{\tau})} \hat{a}_1+\cos \theta \hat{a}_2)$. Again we have got irrelevant phase multipliers in front of the brackets, which can be disregarded. Introducing the notion of phase difference $\phi=\phi_{\rho}-\phi_{\tau}$, we obtain the BS matrix $U_{BS}$ in the form Eq.(\ref{3}). 

Representation of the wave function at the BS output ports in the case of monochromatic beam Eq. (\ref{3}) and non-monochromatic Eq. (\ref{6}) is based on the number of photons in each mode of the electromagnetic field. Since the electromagnetic field is two-mode, an alternative representation of Eqs. (\ref{3}) and (\ref{6}) is possible, which is based not on the number of photons in each mode, but on the angular momentum $L$, where also there are 2 quantum numbers, which are the orbital momentum $l$ and the projection of orbital momentum on a chosen axis $-l \leq m \leq l$ \cite{Campos_1989, Biedenharn_1965}.
For this we can introduce three new operators ${\hat L}_i$ ($i=1,2,3$) which satisfy the commutation rules for angular momentum $[{\hat L}_i,{\hat L}_j]=i\epsilon_{i,j,k}{\hat L}_k]$ ($\epsilon_{i,j,k}$ is a Levi-Civita tensor), where
\begin{eqnarray}
{\hat L}_1=\frac{1}{2}\left({{\hat a_1}^{\dagger}} \hat{a}_2 + {{\hat a_2}^{\dagger}} \hat{a}_1\right),
\nonumber\\
{\hat L}_2=\frac{1}{2i}\left({{\hat a_1}^{\dagger}} \hat{a}_2 - {{\hat a_2}^{\dagger}} \hat{a}_1\right),
\nonumber\\
{\hat L}_3=\frac{1}{2}\left({{\hat a_1}^{\dagger}} \hat{a}_1 - {{\hat a_2}^{\dagger}} \hat{a}_2\right).
\label{11}
\end{eqnarray}
The square, and projection, of the angular momentum are related to the boson number operators via
\begin{eqnarray}
{\hat L}^2=\sum^3_{i=1}{\hat L}^2_i={\hat l}({\hat l}+1),~~ {\hat L}_3={\hat m},
\nonumber\\
{\hat l}=\frac{1}{2}\left({\hat n}_1+{\hat n}_2 \right),~~ {\hat m}=\frac{1}{2}\left({\hat n}_1-{\hat n}_2 \right),
\label{12}
\end{eqnarray}
where ${\hat n}_1, {\hat n}_2$ are operators of the number of particles (photons) in modes 1 and 2 respectively. The BS output ports will be accordingly
\begin{eqnarray}
{\hat L}^{'}_1=\frac{1}{2}\left({{\hat b_1}^{\dagger}} \hat{b}_2 + {{\hat b_2}^{\dagger}} \hat{b}_1\right),
\nonumber\\
{\hat L}^{'}_2=\frac{1}{2i}\left({{\hat b_1}^{\dagger}} \hat{b}_2 - {{\hat b_2}^{\dagger}} \hat{b}_1\right),
\nonumber\\
{\hat L}^{'}_3=\frac{1}{2}\left({{\hat b_1}^{\dagger}} \hat{b}_1 - {{\hat b_2}^{\dagger}} \hat{b}_2\right).
\label{13}
\end{eqnarray}
Thus, it is necessary to introduce a unitary operator ${\hat U}$ that is defined by the expression ${\hat L}^{'}_i={\hat U}{\hat L}_i {{\hat U}^{\dagger}}$. Since this operator is responsible for the transformation of the angular momentum, which is responsible for the symmetry during rotations, the general form of such an operator should be of the form ${\hat U}=e^{-i\Phi {\hat L}_3}e^{-i\Theta {\hat L}_2} e^{-i\Psi {\hat L}_3}$, where the parameters ($\Phi, \Theta, \Psi $) are quantum-mechanical counterparts to the classical Euler angles. Each angle has its own physical meaning, which we will try to define next. Using the Baker-Campbell-Hausdorff formula you can get \cite{Campos_1989}
\begin{eqnarray}
\begin{pmatrix}
  \hat{b}_1\\\
  \hat{b}_2
\end{pmatrix}=
U
\begin{pmatrix}
  \hat{a}_1\\\
  \hat{a}_2
\end{pmatrix},
~~~
U=\begin{pmatrix}
\cos (\theta/2) e^{i(\Psi+\Phi)/2} & \sin (\theta/2) e^{i(\Psi-\Phi)/2}\\\
-\sin (\theta/2) e^{-i(\Psi-\Phi)/2}& \cos \theta e^{-i(\Psi+\Phi)/2}.
\end{pmatrix} 
\label{14}
\end{eqnarray}
Let's compare the obtained Eq. (\ref{14}) c Eq. (\ref{10}), then we get (up to an insignificant phase)
\begin{eqnarray}
\Theta=2\theta,~~~ \Phi+\Psi=2\phi_{\tau}, ~~~\Psi-\Phi = 2 \phi_{\rho}.
\label{15}
\end{eqnarray}
The result is
\begin{eqnarray}
{\hat U}=e^{-i(\phi_{\tau}-\phi_{\rho}) {\hat L}_1}e^{-2i arccos\sqrt{1-R}~ {\hat L}_2} e^{-i(\phi_{\tau}+\phi_{\rho}) {\hat L}_3}.
\label{16}
\end{eqnarray}
If we introduce step-up and step-down operators ${\hat L}_{\pm}={\hat L}_{1} \pm i{\hat L}_{2}$ (${\hat L}_{+}={{\hat a}^{\dagger}}_1 {\hat a}_2,~{\hat L}_{-}={{\hat a}^{\dagger}}_2 {\hat a}_1$), then the operator ${\hat U}$ can be represented as
\begin{eqnarray}
{\hat U}={{\hat D}^{\dagger}} e^{-2i\phi_{\tau} {\hat L}_3} ,~~{\hat D}=e^{\zeta {\hat L}_{+}-\zeta^{*} {\hat L}_{-}},
\label{17}
\end{eqnarray}
where $\zeta=arccos\sqrt{1-R}~ e^{i \phi}$ (where $\phi=\phi_{\rho}-\phi_{\tau}$), and the operators ${\hat L}_{\pm}$ have known properties ${\hat L}_{\pm} |l,m \rangle = \{ l(l+1)-m(m\pm 1){\}}^{1/2} |l,m \pm 1 \rangle$. Eq.(\ref{17}) can be further simplified by ignoring the irrelevant phase factor which will give the operator $e^{-2i\phi_{\tau} {\hat L}_3}$, since $e^{-2i\phi_{\tau} {\hat L}_3}|l,m \rangle=e^{-2i\phi_{\tau} m} |l,m \rangle$, then we get that ${\hat U}={{\hat D}^{\dagger}}$.  As a result, using Eq.(\ref{17}) we can find the wave function on the BS output ports $|\Psi_{out}\rangle={\hat U}^{\dagger}|\Psi_{in}\rangle$. 

\begin{figure}[!h]
\center{\includegraphics[angle=0, width=0.9\textwidth, keepaspectratio]{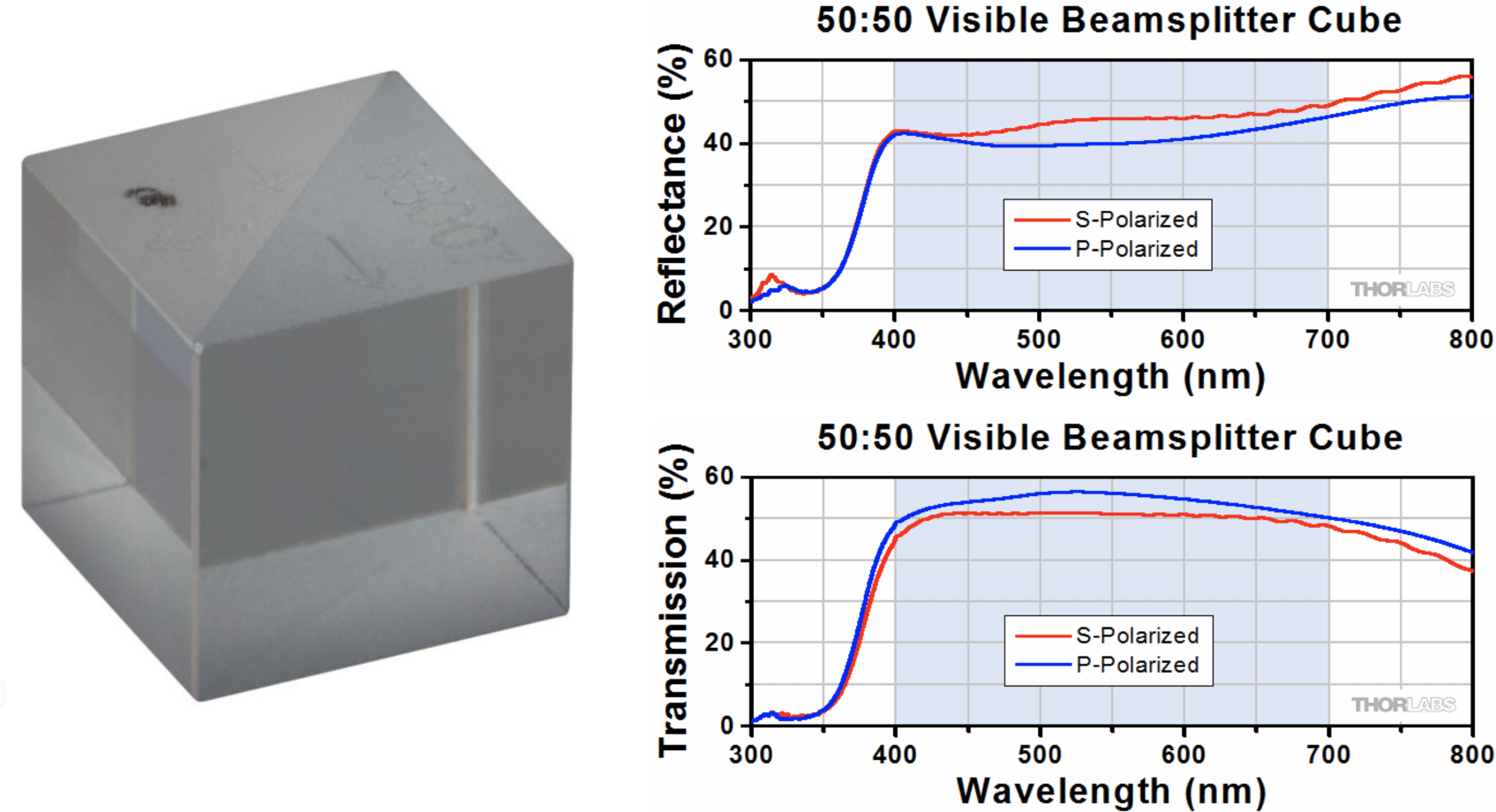}} 
\caption[fig_2]{A sample of an non-polarizing cubic beam splitter (50:50), with an operating wavelength range of (400 - 700) nm, BS dimensions of 5 mm, is presented. The figures on the right show the dependence of reflection coefficient $R$ and transmittance $T$ as a function of wavelength $\lambda$. BS fabrication accuracy: $T=47\pm 10\%$, $R=47\pm 10\%$, $R+T>90\%$ \cite{BS_Cub}.}
\label{fig_2}
\end{figure}

This result is quite interesting scientifically because without solving the Schrodinger equation we can find the wave function $|\Psi_{out}\rangle$ at the BS output ports knowing $R$ and phase shift $\phi$. Moreover, we are not interested in the type of BS and this solution is suitable for all two-port linear BS. Of course, with such a general consideration, the ``nature'' of the appearance of coefficients $R$ and $\phi$ in quantum optics is not clear, since they have been introduced as some constants having the properties of these coefficients.  In spite of this, the ``nature'' of the appearance of these coefficients is well studied in classical optics for various types of BS. For different types of BS, these coefficients can be calculated using classical optics, where the dependence $R=R(\lambda)$ (or in the frequency representation $R=R(\omega)$) appears.    It should be added that the phase shift $\phi$ does not affect the intensity at the BS output ports, so it may not be considered. For example, for a cubic BS, which is often used in quantum optics experiments, the dependence can be represented in Fig. \ref{fig_2}.
It should be added that to find the wave function $\Psi_{out}$, taking into account the non-monochromaticity of photons, using Eq. (\ref{6}) there is a rather interesting case, which is not discussed in the textbook and review literature, but which is fundamentally important for waveguide BS. Non-monochromatic photons are known to have some frequency distribution $\phi (\omega_1, \omega_2)$, i.e. the frequencies $\omega_1, \omega_2$ are ``blurred'' and the amount of blurring is determined by the variance $\sigma$. It should be added that the optical sources (including single photon sources) that are commonly used are such that $\sigma \ll \omega$.  In order to find $\Psi_{out}$ it is necessary to integrate over the frequencies $\omega_1, \omega_2$, see Eq. (\ref{6}). With this integration, the coefficients $R$ and $T$ can be assumed to be constant, since $\{R(\omega\pm\sigma)-R(\omega)\} \to 0$. This is well seen from Fig. (\ref{fig_2}), where by changing $\lambda$ by a small value $\Delta \lambda \ll \lambda$, the coefficients $R$ and $T$ will also change by a small value (i.e. $\Delta R \ll R$), whose change can be neglected. So by choosing constant values of $R$ and $T$ we do not make a mistake when $R$ and $T$ are smooth (no jumps) functions with respect to frequencies. Of course, it is reasonable to use such smooth functions and at first sight there can be no other options. As will be shown in the next section, that such functions appear in the case of waveguide BS and this will significantly affect the statistical properties of photons at the BS output ports. Moreover, quantum entanglement of photons will strongly depend on it, which certainly changes the relation to the waveguide BS as a source of quantum entangled photons. Thus, in this review, BSs where $R$ and $\phi$ are assumed to be constant will be considered, which will be called ``conventional'' beam splitters, and where $R$ and $\phi$ are not constant, called frequency-dependent BSs. So far, there is information in the literature about only one frequency-dependent type of BS, which is the waveguide BS.
\subsection{``Conventional'' waveguide beam splitter}
\begin{figure}[!h]
\center{\includegraphics[angle=0, width=0.9\textwidth, keepaspectratio]{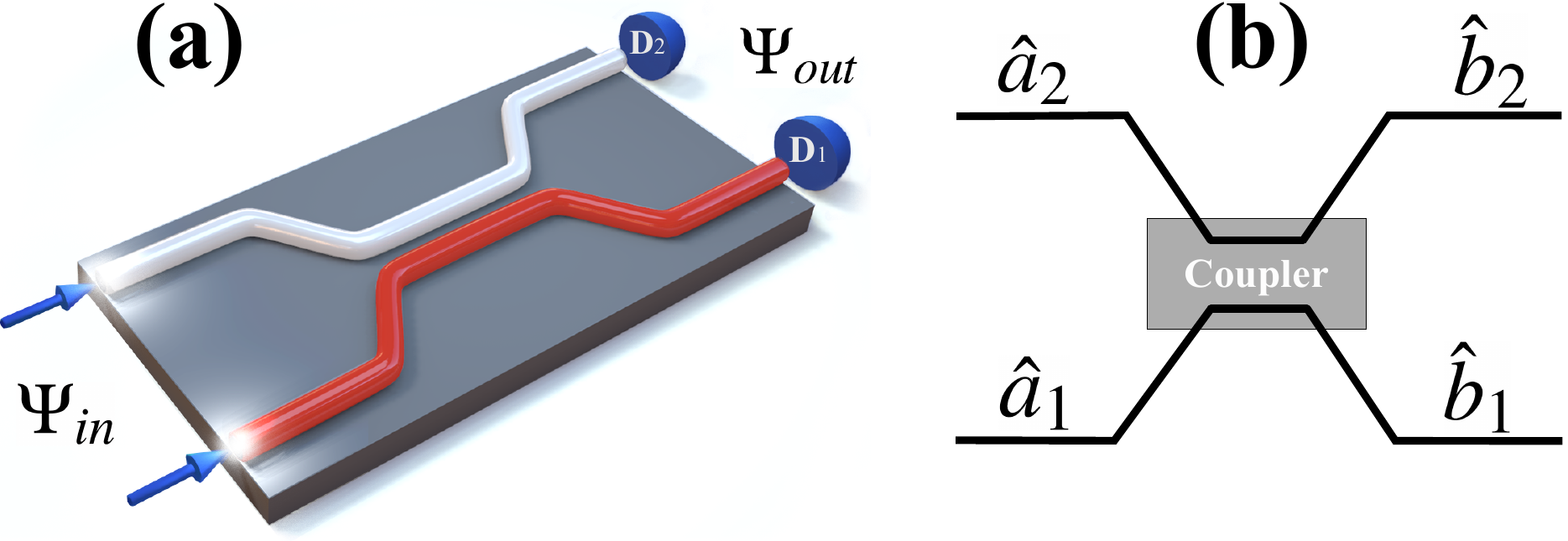}} 
\caption[fig_3]{3D representation of coupled waveguides, where $ \Psi_{in}$ and $ \Psi_{out}$ are the input and output (recorded by $D_1$ and $D_2$ detectors) wave function of photons, respectively. (b): Schematic 2D representation of such a BS, see \cite{Makarov_SR2_2021}.}
\label{fig_3}
\end{figure}

It should be added that all the expressions obtained above for ``conventional'' BS, are suitable for waveguide BS. Nevertheless, we will show how in theory two coupled waveguides begin to exhibit BS properties. 
Consider two waveguides connected (coupled) to each other in a certain area of space, see Fig.\ref{fig_3}. 
The theory of coupled waveguides is now quite well developed \cite{Huang_1994}. This theory is based on coupling coefficients between neighbouring waveguides. It is generally believed that by setting these coupling coefficients one can regulate the photon states at the output ports of the waveguides. In fact, this theory is applicable for identical and monochromatic photons incident on incoming ports of waveguides. Or as an alternative representation: coupling coefficients depend on frequencies but how they do not matter since these coefficients are given by constants. As it has been shown recently \cite{Makarov_OL_2020,Makarov_SR_2020,Makarov_PRE_2020,Makarov_SR1_2021,Makarov_SR2_2021}, on the basis of exact quantum-mechanical consideration, that states of photons at output ports of waveguides have strong dependence on their frequencies and to set coupling coefficients by constant values is not always correct and it can carry large errors. Waveguide theory (more exactly it is a BS theory) under exact quantum-mechanical consideration in the particular case of monochromatic and identical photons turns into a waveguide theory with constant coefficients. Let us first introduce here the theory of waveguide BS based on constant coupling coefficients.

It is well known that for coupled waveguides one can compose a system of two equations \cite{Huang_1994}
\begin{eqnarray}
\frac{d \hat{a}_1}{dz}=-i(\beta_1+K_{11})\hat{a}_1-iK_{12}\hat{a}_2,
\nonumber\\
\frac{d \hat{a}_2}{dz}=-i(\beta_2+K_{22})\hat{a}_2-iK_{21}\hat{a}_1,
\label{18}
\end{eqnarray}
where $\beta_1, \beta_2$ these are the propagation constants for modes 1 and 2 respectively;  $K_{12},K_{21}$ and $K_{11},K_{22}$ are the mutual and the self coupling coefficients, respectively; $z$ is the coordinate at which the field characteristics are determined. Moreover, if the coupled waveguide system is lossless, self-consistency requires that the Eq. (\ref{18}) satisfy the power conservation law, i.e. at the input and output ports the number of photons does not change. Because Eq. (\ref{18}) is independent of the initial condition, the coupling coefficients obey the following relations $K_{12}=K^{*}_{21}=\kappa$ and $K_{11},K_{22}$ have to be real. The solution of Eq.(\ref{18}) is conveniently found in the form $\hat{a}_i \to \hat{a}_i e^{-i/2(\beta_1+\beta_2+K_{11}+K_{22})z}$. In this form, the coordinate $z$ will have the meaning of the waveguides coupling region length. Using this form we obtain
\begin{eqnarray}
\frac{d \hat{a}_1}{dz}=-i\delta \hat{a}_1-i \kappa \hat{a}_2,
\nonumber\\
\frac{d \hat{a}_2}{dz}=+i\delta \hat{a}_2-i \kappa \hat{a}_1,
\label{19}
\end{eqnarray}
where $\delta=1/2(\beta_1 - \beta_2+K_{11} - K_{22})$. In this form, solve Eq. (\ref{19}) is simple enough, eg \cite{Huang_1994}. If we adopt new notation in the form $\hat{a}_i(0)=\hat{a}_i$ and $\hat{a}_i(z)=\hat{b}_i$, then we get
\begin{eqnarray}
\begin{pmatrix}
  \hat{b}_1\\\
  \hat{b}_2
\end{pmatrix}=
\begin{pmatrix}
u_{11}& u_{12}\\\
u_{21}& u_{22}
\end{pmatrix}
\begin{pmatrix}
  \hat{a}_1\\\
  \hat{a}_2
\end{pmatrix},
\label{20}
\end{eqnarray}
where $u_{11}=u^{*}_{22}=\cos(S z)-i \cos\eta \sin(Sz)$ and $t_{12}=t_{21}=-i\sin\eta\sin(S z)$, and the parameters $S=\sqrt{\delta^2+\kappa^2},~ \tan\eta=\kappa/\delta$. On ports 1 and 2 output, the probabilities of detecting photons would be, respectively $P_1=\cos^2(S z)+\cos^2\eta \sin^2(Sz), P_2=\sin^2\eta\sin^2(S z)$. The maximum probability value on the 2nd output port is $(P_2)_{max}=\sin^2\eta$. If photons propagate in waveguides without loss and the number of photons is conserved, then $(P_2)_{max}=1$, which means $\eta=\pm\pi/2$.
In this case we will get
\begin{eqnarray}
\begin{pmatrix}
  \hat{b}_1\\\
  \hat{b}_2
\end{pmatrix}=
\begin{pmatrix}
\cos(\kappa z)& \pm i \sin(\kappa z)\\\
\pm i \sin(\kappa z)& \cos(\kappa z)
\end{pmatrix}
\begin{pmatrix}
  \hat{a}_1\\\
  \hat{a}_2
\end{pmatrix}.
\label{21}
\end{eqnarray}
If we put $\sqrt{T}=\cos(\kappa z)$, then the result is in Eq. (\ref{21}) is the same as the beam splitter, see Eq.(\ref{4}). In this consideration, we have not only shown that coupled waveguides have BS property, but also found in simple form reflections $R=\sin^2(\kappa z)$ \cite{Bromberg_2009}.

Obviously, the case considered here is based on coupling coefficients, but does not reveal the ``nature'' of the appearance of these coefficients. Moreover, it is quite obvious, even by analogy with ``conventional'' BS, that there must be a dependence of reflection coefficients $R$ on frequencies of the incident field, i.e. $R=R(\omega)$. In order to find this dependence it is necessary to consider a quantum mechanical system in which two quantized modes of electromagnetic field interact with atoms of matter in coupled waveguides. Such a consideration was made recently in the papers \cite{Makarov_OL_2020,Makarov_SR_2020,Makarov_PRE_2020,Makarov_SR1_2021,Makarov_SR2_2021} and the dependence not only of $R=R(\omega)$, but also of the phase shift $\phi=\phi(\omega)$ was found. Let us now turn to this theory.

\subsection{Frequency-dependent waveguide beam splitter}
Consider two coupled waveguides, along each of which photons propagate. These photons interact with atomic electrons. Let us represent the electromagnetic field of photons through the transverse vector potential $ {\bf A} $ in the Coulomb gauge $ div {\bf A} = 0 $ \cite{Mandel_1995,Agarwal_2013}, then the Hamiltonian of such system will be
\begin{eqnarray}
\left\lbrace {\hat H}_{1}+{\hat H}_{2}+\frac{1}{2}\sum_a\left({\hat{\bf p}}_a+\frac{1}{c}\hat{\bf A}_a\right)^2 + \sum_a U({\bf r}_a)\right\rbrace \Psi = i\frac{\partial \Psi}{\partial t}, 
\label{22}
\end{eqnarray}
where $ {\hat H}_{i} =\omega_i\hat{a_i}^{+}\hat {a_i} $ is the Hamiltonian operator for the first ($ i=1 $) and second ($ i = 2 $) modes ($\omega_i $ is the frequency and $\hat{a_i} $ is the photon annihilation operator in mode number $ i $); $ U ({\bf r}_a) $ is the atomic potential acting on the electron with number $ a $; $ {\hat{\bf p}}_a $ is the momentum operator of the electron with number $ a $; $ \hat{\bf A}_a = \hat{\bf A}_{1, a} + \hat{\bf A}_{2,a} $, where $ \hat {\bf A}_{i, a} = \sqrt{\frac{2 \pi c^2}{\omega_i V}}{\bf u}_i(\hat{a_i}^{+}+\hat{a_i}) $ is the vector potential in dipole approximation for mode $ i $, acting on an electron with number $ a $ ($ c $ is the speed of light, $ V $ is the modal volume, i.e. it is the Fourier transform of the vector potential ${\bf A}$ enclosed in a finite volume $V$, $ {\bf u}_i $ is the polarization of the mode numbered $ i $) \cite{Mandel_1995, Agarwal_2013}; the sum $\sum_a $ in Eq.(\ref{22}) is satisfied over all electrons of the two coupled waveguides interacting with photons.  As a result, the Hamiltonian of equation (\ref{22}) will be
\begin{eqnarray}
{\hat H}= \sum^{2}_{i=1}\left\lbrace \omega_i\hat{a_i}^{+}\hat {a_i}+\overline{N}\frac{\beta^2_i}{2}q^2_i+\beta_i q_i{\bf u}_i\sum_a {\hat{\bf p}_a}\right\rbrace+
\nonumber\\
\overline{N}\beta_1 \beta_2 q_1 q_2 {\bf u}_1 {\bf u}_2+\sum_a\frac{{\hat{\bf p}}^2_a}{2}+\sum_a U({\bf r}_a), 
\label{23}
\end{eqnarray}
where $\beta_i = \sqrt{\frac{4 \pi}{\omega_i V}} $, and the quantity $ \overline{N}= \sum_a$ is the number of electrons involved in the interaction with photons in two coupled waveguides.
As can be seen, equation (\ref{23}) corresponds to the equation for coupled harmonic oscillators interacting with electrons of a polyatomic system. A similar system was considered in \cite{Makarov_PRE_2020}, we obtain 
\begin{eqnarray}
{\hat H}= \sum^{2}_{i=1} \left\lbrace \frac{\sqrt{A_i}}{2} \left( {\hat P}^2_i+y^2_i\right)+{\bf D}_i y_i\sum_a {\hat{\bf p}_a}\right\rbrace +
\sum_a \frac{{\hat{\bf p}}^2_a}{2}+\sum_a U({\bf r}_a) , 
\label{24}
\end{eqnarray}
where $ y_1 =  A^{1/4}_1\left( q_1/\sqrt{\omega_1} \cos \alpha-q_2/\sqrt{\omega_2} \sin \alpha\right)$ and $y_2 = A^{1/4}_2\left(q_1/\sqrt{\omega_1} \sin \alpha + q_2/\sqrt{\omega_2} \cos \alpha \right)$ is new variables; ${\hat P}_i=-i\partial/\partial y_i$;  ${\bf D}_1=\beta_1 A^{-1/4}_1\sqrt{\omega_1}{\bf u}_1\cos\alpha - \beta_2  A^{-1/4}_1\sqrt{\omega_2}{\bf u}_2 \sin\alpha$; ${\bf D}_2=\beta_1 A^{-1/4}_2\sqrt{\omega_1}{\bf u}_1\sin\alpha+\beta_2A^{-1/4}_2 \sqrt{\omega_2}{\bf u}_2 \cos\alpha$. The study \cite{Makarov_2018_PRE} showed that $\tan(2\alpha)=C/(B_2-B_1)$, where $C=2\overline{N} \beta_1 \beta_2 \sqrt{\omega_1 \omega_2} {\bf u}_1 {\bf u}_2$; $B_i=(\omega_i+\overline{N}\beta^2_i)\omega_i$ and $A_i=B_i+(-1)^i C/2 \tan\alpha$.
Obviously, the value of $ \beta_i $ is very small in the case of single-photon interaction, see, eg \cite{Tey_2008}, where $ \beta^2/\omega \ll 1 $, even in the case of strong focusing. In this case, the quantities $ {\bf D}_1 $ and $ {\bf D}_2 $ are negligible. This is an obvious fact, since these quantities are responsible for various inelastic transitions of electrons in an atom under the action of photons, which are usually negligible in lossless BS. As a result, the dynamics of two photons in BS will be described by the wave function
\begin{eqnarray}
|\Psi(t_{BS})\rangle =e^{-i {\hat H_{BS}}t_{BS}}|\Psi(0)\rangle  , ~~~ \hat H_{BS} = \sum^{2}_{i=1} \frac{\sqrt{A_i}}{2}\left\lbrace {\hat P}^2_i+y^2_i\right\rbrace ,
\label{25}
\end{eqnarray}
where $ t_{BS} $ is the photon interaction time in BS and $ |\Psi(0)\rangle $ is the initial state of the photons before entering the BS. It should be added that the representation of the solution to the general Eq. (\ref{22}) in the form Eq. (\ref{25}) allows us to determine in an analytical form the evolution of the operators $ \hat{a_1} $ and $ \hat{a}_2 $ in Eq.(\ref{3}). Often, to calculate the required quantities, one needs electric field operators $ {\hat E}_1(t_1) $ and $ {\hat E}_2(t_2) $ at time instants $ t_1 $ and $ t_2 $ on the first and second detectors, respectively. To do this, we need to find the evolution (in BS, as well as from BS to detectors) of the operators $ {\hat E}_{01}(0) $ and ${\hat E}_{02}(0) $ for modes 1 and 2, respectively
\begin{eqnarray}
{\hat E}_1(t_1)=e^{i {\hat H_0}t_1}e^{i {\hat H_{BS}}t_{BS}}{\hat E}_{01}(0)e^{-i {\hat H_{BS}}t_{BS}} e^{-i {\hat H_0}t_1},
\nonumber\\
{\hat E}_2(t_2)=e^{i {\hat H_0}t_2}e^{i {\hat H_{BS}}t_{BS}}{\hat E}_{02}(0)e^{-i {\hat H_{BS}}t_{BS}} e^{-i {\hat H_0}t_2},
\label{26}
\end{eqnarray}
where $ {\hat H_0} = \sum^{2}_{i = 1} \omega_i\hat{a_i}^{\dagger}\hat {a_i}$ is the Hamiltonian of photons outside of BS. Because $ {\hat E}_{01}(0) \propto{\hat a}_1 $, and $ {\hat E}_{02}(0) \propto {\hat a}_2 $ (see, eg \cite{Mandel_1995,Agarwal_2013}), it is more convenient to consider not the electric field operators, but the photon creation and annihilation operators before entering BS ($ {\hat a}_1 $ and $ {\hat a}_2 $) and on the detectors ($ {\hat b}_1 $ and $ {\hat b}_2 $). To this end, we replace $ {\hat E}_{01}(0)\to{\hat a}_1, {\hat E}_{02}(0)\to{\hat a}_2 $ and $ {\hat E}_{1} (t_1) \to{\hat b}_1(t_1), {\hat E}_{2}(t_2) \to{\hat b}_2(t_2) $. Including into account the time delay $ \delta \tau $ for the spatial displacement of BS from the equilibrium position and the time delay $ \tau $ between 1 and 2 mods (see Fig. \ref{fig_4}), we get (see supplementary materials in \cite{Makarov_OL_2020})
\begin{eqnarray}
{\hat b}_1(t_1)=\sqrt{T}  e^{-i\omega_1 (t_1-\tau)}{\hat a}_1+e^{i\phi}\sqrt{R} e^{-i\omega_2 (t_1+\delta \tau/2)}{\hat a}_2,
\nonumber\\
{\hat b}_2(t_2)=\sqrt{T} e^{-i\omega_2 t_2}{\hat a}_2 -e^{-i\phi}\sqrt{R} e^{-i\omega_1 (t_2-\delta \tau/2-\tau)}{\hat a}_1,
\label{27}
\end{eqnarray}
where
\begin{eqnarray}
R=\frac{\sin^2\left( \Omega t_{BS}/2 \sqrt{1+\varepsilon^2}  \right) }{(1+\varepsilon^2)};~\cos\phi=-\varepsilon \sqrt{\frac{R}{T}},
\nonumber\\
T=1-R, ~~\Omega=\frac{8\pi \overline{N} {\bf u}_1 {\bf u}_2  }{(\omega_1+\omega_2)V};~\varepsilon=\frac{\omega_2 -\omega_1}{\Omega}.
\label{28}
\end{eqnarray}
\begin{figure}[!h]
\center{\includegraphics[angle=0, width=0.75\textwidth, keepaspectratio]{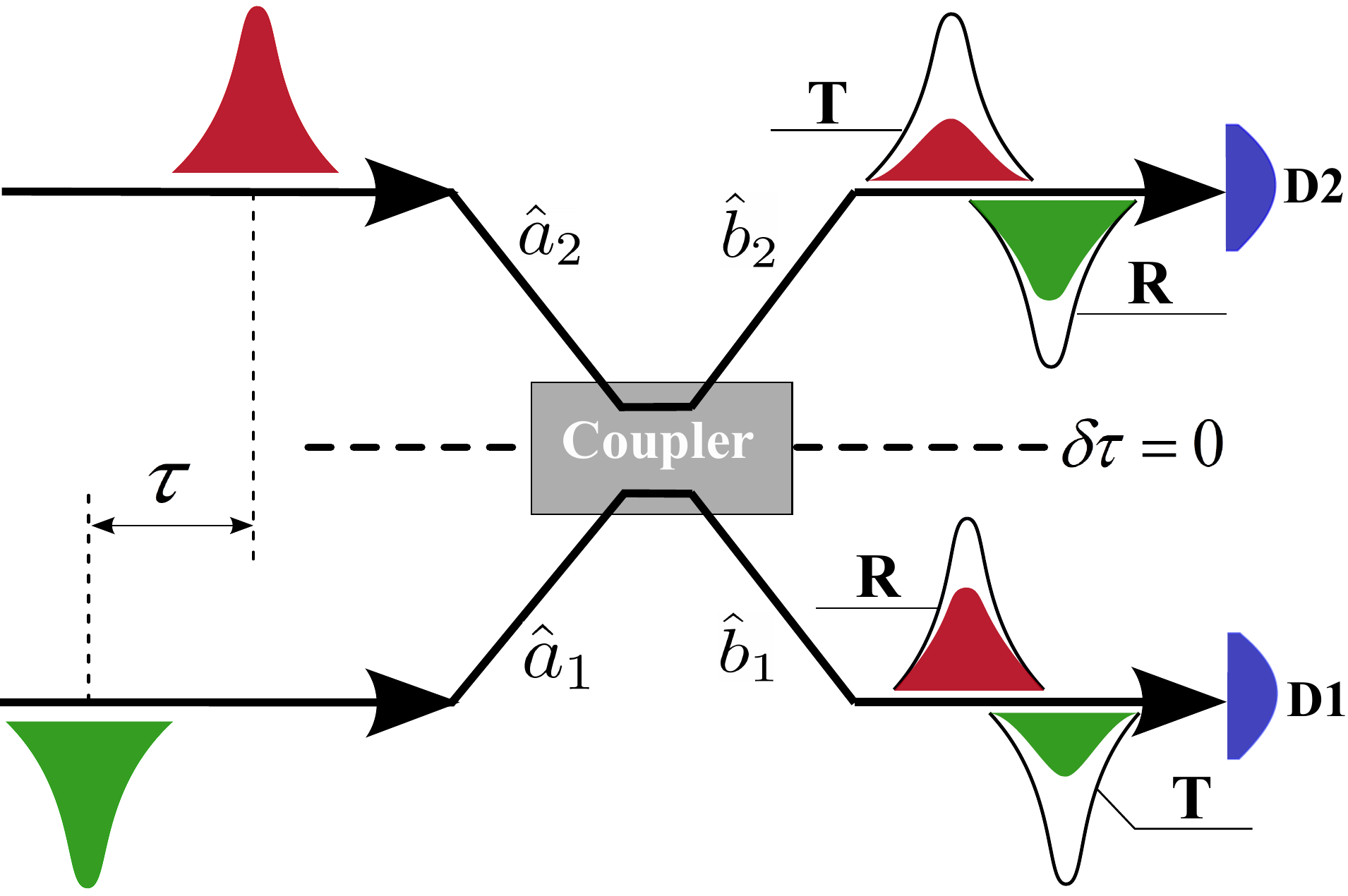}} 
\caption[fig_4]{Schematic representation of a waveguide BS, where $D_1, D_2$ are first and second detectors respectively; $\tau$ is the time delay between modes 1 and 2, and $\delta \tau$ is the time delay caused by the spatial shift of the BS from its equilibrium position.}
\label{fig_4}
\end{figure}
From (\ref{27}) it can be seen that the matrix BS that is $ U_ {BS} $ completely corresponds to the matrix Eq.(\ref{3}) (needless to say, for $ t_1 = t_2 = t $ and for $ \delta \tau = \tau = 0 $). In addition, $ T $ and $ R $ are symmetric, i.e. if we change the first to the second mode $ \omega_1 \to \omega_2, {\bf u}_1 \to{\bf u} _2 $ and vice versa, then, as anticipated, the coefficients will not alter. If we assume that photons are completely identical and monochromatic then $\varepsilon=0$ (in reality this does not happen) and we obtain the ``conventional'' waveguide BS described above (at $\kappa = \Omega/(2v)$, $z= v t_{BS}$, where $v$ is the wave speed in the waveguide). This means that this particular case will fully coincide with Eq.(\ref{21}). It can be seen that the theory presented here is general, where the frequency dependencies of the reflection coefficient $R$ and phase shift $\phi$ are found.

From Eq. (\ref{28}) it can be seen that there is a new quantity characterizing BS and it is $\Omega$. It can be qualitatively estimated if we take into account that most of quantum effects are observed at $\omega_2-\omega_1\ll \omega_1, \omega_2$. We will then be interested in the frequencies at $\omega_2\approx \omega_1$; then $\Omega=4\pi n/\omega_0 {\bf u}_1 {\bf u}_2$, where $\omega_0=(\omega_2+\omega_1)/2$, $n=\overline{N}/V$. In this representation $\omega_0$ can be thought of as a constant and $n$ as the effective concentration. Moreover, even if the condition $\omega_2-\omega_1\ll \omega_1 , \omega_2$ is not met, and $\omega_2-\omega_1\sim \omega_1 , \omega_2$, then $\omega_0$ can be considered approximately constant when integrated over frequencies in Eq. (\ref{6}). Note that $n$ is a characteristic of the waveguide BS, since it determines the coupling of effectively interacting $\overline{N}$ electrons to photons in the volume $V$. For example, if we consider the BS as a solid in which photons interact in a volume $V$, then $n$ is the concentration of electrons in the solid. Since it is known that the concentration of atoms in a solid does not vary much, we get $\Omega\sim (10^{14}-10^{17})rad/s$ for the optical frequencies ($\omega_0\approx 10^{15} rad/s$) in this case (if of course ${\bf u}_1 {\bf u}_2\sim 1$). It should be added that for this type of BS, the frequency $\Omega$ can be represented by the known value for plasma frequency $\omega_p$, then $\Omega={\bf u}_1 {\bf u}_2 \omega^2_p/\omega_0$.  In a waveguide BS, the number of effectively interacting electrons $\overline{N}$ in the volume $V$ is smaller than in a solid, hence $\Omega$ will be smaller. In waveguide BS one can adjust $\Omega$ by changing the effective interaction between photons. 

From Eq. (\ref{28}) we can see that $R=R(\omega_1, \omega_2)$ and $\phi=\phi(\omega_1, \omega_2)$, i.e. depends on frequencies of 1 and 2 modes. This is the main difference between waveguide BS and the other type of waveguide, where the reflection coefficient $R$ depends on the frequency of one mode. This result causes the reflection coefficient $R$ to become ``sensitive'' to changes in any of the frequencies in the mode. Considering that the $\Omega$ parameter is quite small, the dependence of the reflection coefficient $R$ on the frequency can have a dramatic change. In other words, such a dependence has a resonance character at $(\omega_2-\omega_1)\sim \Omega$. Indeed, by changing $R(\omega_1\pm \sigma_1, \omega_2\pm \sigma_2)$ the reflection coefficient value will change greatly at $\sigma_1\sim \Omega, \sigma_2\sim \Omega$, where $\sigma$ is the variance that was described above. Otherwise, if $\sigma_1\ll \Omega, \sigma_2\ll \Omega$, then photons in the modes can be considered monochromatic and the resonance dependence $R(\omega_1, \omega_2)$ disappears, which means in this case the reflection coefficient $R$ can be considered constant and such waveguide BS becomes ``conventional''\cite{Makarov_SR1_2021}.
\section{Quantum entanglement of photons on a beam splitter}
As is well known, quantum entanglement is a phenomenon where the quantum state of each group of particles cannot be described independently of the state of the others, including when the particles are separated by a large distance. Quantum entanglement is now one of the most important research topics \cite{Horodecki_2009}.  In particular, quantum communication protocols such as quantum cryptography \cite{Ekert_1991}, quantum dense coding \cite{Bennett_1992}, quantum computing algorithms \cite{Shor_1995} and quantum state teleportation \cite{Aspect_1981, Samuel_1998} can be explained by entangled states. One of the promising prospects for using quantum entanglement is to build a quantum computer that will surpass modern classical computers by many orders of magnitude. Moreover, the beam splitter could be used as a source of quantum entangled photons and be the basic building block of a quantum computer \cite{Knill_2001}.  Without going into details of this topic, it is worth saying that a rather serious problem is the calculation of quantum entanglement for multiparticle systems, where different measures of quantum entanglement \cite{Horodecki_2009}, such as \textit{Concurrence}, \textit{Negativity}, etc., are used. For calculations of two-component systems, the quantum entanglement calculation is simpler and is based on well-known measures of quantum entanglement - these are \textit{von Neumann entropy}, \textit{Schmidt parameter}, etc. \cite{Horodecki_2009}.

It is known that by Schmidt's theorem \cite{Ekert_1995,Grobe_1994} the wave function $|\Psi(t)\rangle $ of interacting systems 1 and 2 can be decomposed as $|\Psi(t)\rangle = \sum_{k}\sqrt{\lambda_k (t)}|u_{k}(x_1, t)\rangle |v_{k}(x_2,t)\rangle$, where $|u_{k}(x_1,t)\rangle$ is the pure state wave function of system 1 and $|v_{k}(x_2,t)\rangle$ is the pure state wave function of system 2. Where $\lambda_k$ is the Schmidt mode, which is the eigenvalue of the reduced density matrix,  i.e. $\rho_1(x_1,x^{'}_1,t)=\sum_{k}\lambda_k(t) u_{k}(x_1,t)u^{*}_{k}(x^{'}_{1},t) $ or $\rho_2(x_2,x^{'}_{2},t)=\sum_{k}\lambda_k(t) v_{k}(x_2,t)v^{*}_{k}(x^{'}_{2},t) $. If we find the Schmidt mode $\lambda_k$, we can calculate the quantum entanglement of the system. To do this, we can use various measures of quantum entanglement, such as the Schmidt parameter \cite{Ekert_1995,Grobe_1994} $K=\left(\sum_{k}\lambda^2_k \right)^{-1} $ or von Neumann entropy \cite{Bennett_1996,Casini_2009} $S_N=-\sum_{k} \lambda_k \ln \left(\lambda_k \right) $. The main difficulty in calculating quantum entanglement is to find $\lambda_k $ of the system in question.
\subsection{Quantum entanglement on a ``conventional'' beam splitter}
\begin{figure}[h]
\includegraphics[angle=0,width=0.9\textwidth, keepaspectratio]{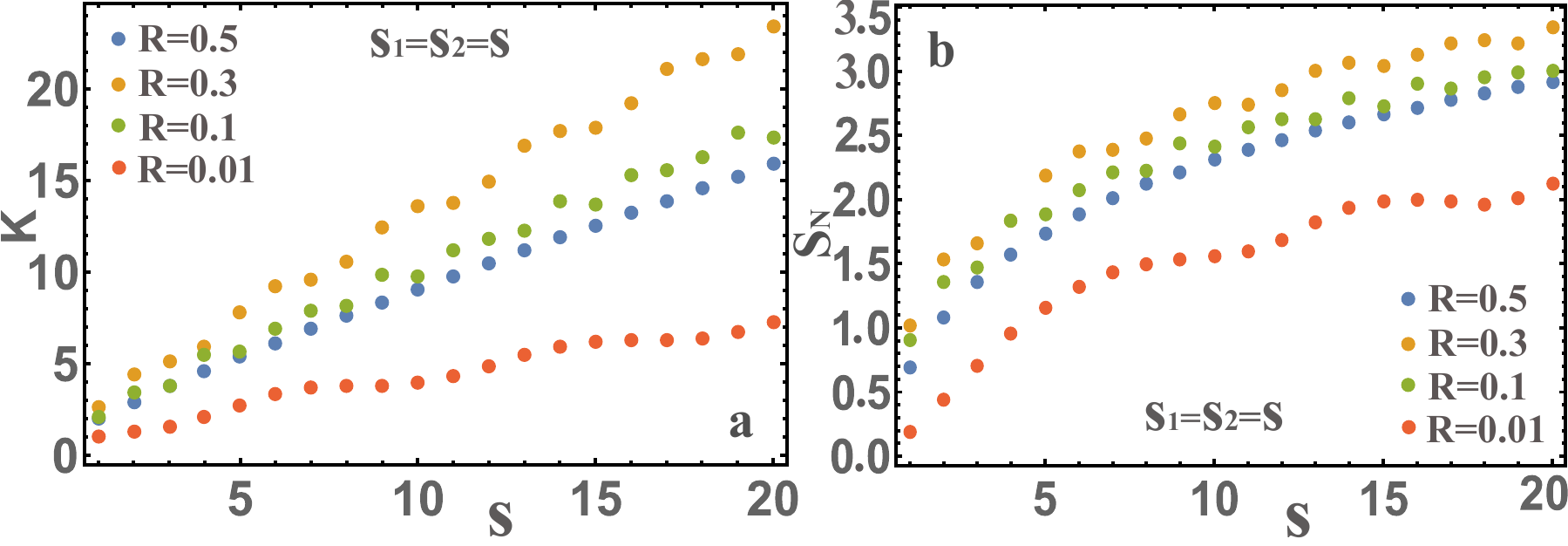}
\caption[fig_5]{The dependence of the von Neumann entropy $S_N$ in Figures (a) and (b), and the Schmidt parameter in Figures (c) and (d) as a function of $R$ are presented. The figures show the dependencies ($s_1, s_2$) for different values of the number of photons in 1 and 2 modes, respectively. For example, when $s_1=1$ and $s_2=6$, notations (1,6) are introduced.}
\label{fig_5}
\end{figure}

Quantum entanglement on ``conventional'' BS is quite well studied and presented in various works, see eg \cite{Kim_2002,Ou_2007,Makarov_PRE_2020,Chen_2021,Jiang_2013,Berrada_2009,Wang_2002}. Let us present here the simplest derivation of the quantum entanglement of the two-mode electromagnetic field on BS.  The initial states of photons $|\Psi_{in}\rangle$ in BS ports 1 and 2 will be chosen as Fock states. Consideration of other initial states can be found in many works, but the basis of all works are Fock states.   In this case, from Eq.(\ref{2}) we obtain 
\begin{eqnarray}
|\Psi_{in}\rangle=\frac{1}{\sqrt{s_1! s_2!}}{{\hat a_1}^{\dagger}}{}^{s_1} {{\hat a_2}^{\dagger}}{}^{s_2}|0\rangle, ~~|\Psi_{out}\rangle=\frac{1}{\sqrt{s_1! s_2!}}{{\hat b_1}^{\dagger}}{}^{s_1} {{\hat b_2}^{\dagger}}{}^{s_2}|0\rangle .
\label{29}
\end{eqnarray}
To find $\Psi_{out}$ you can use Eq.(\ref{17}), as was done in \cite{Kim_2002} using the \cite{Campos_1989} approach. This approach is not the easiest to calculate and analyse quantum entanglement. The simplest approach has been proposed with \cite{Makarov_PRE_2020}, using the \cite{Makarov_2017_adf} approach. As a result, it was obtained $\lambda_k (R)=\left|c_{k,s_1+s_2-k}\right|^2 $, where
\begin{eqnarray}
c_{k,p}=\sum^{s_1+s_2}_{n=0}A^{s_1,s_2}_{n,s_1+s_2-n}A^{*{k,p}}_{n,s_1+s_2-n}e^{-2in ~ {\arccos\left( \sqrt{1-R} \sin\phi \right)}} ,
\nonumber\\
A^{k,p}_{n,m}=\frac{\mu^{k+n}\sqrt{m!n!}}{(1+\mu^2)^{\frac{n+m}{2}}\sqrt{k!p!}}P^{(-(1+m+n), m-k)}_{n}\left(-\frac{2+\mu^2}{\mu^2} \right),
\nonumber\\
\mu =\sqrt{1+\frac{1-R}{R}\cos^2\phi}-\cos\phi\sqrt{\frac{1-R}{R}},
\label{30}
\end{eqnarray}
where $P^{\alpha, \beta}_{\gamma}(x)$ are Jacobi polynomials, $k$ and $p$ are  the number of photons at the output ports 1 and 2 of the BS respectively, with $k+p=s_1+s_2$. In the Eq.(\ref{30}), no matter what value $\phi\in (0,\pi/2)$ we choose, the value $\lambda_k(R)$ will not depend on $\phi$. This amazing property of the Eq. (\ref{30}) can be used depending on the tasks under consideration. For example, if we choose $\phi=0$, then when summing by $n$ in (\ref{30}), the exponent is replaced by $(-1)^n$. if we choose $\phi=\pi/2$, then all the dependence on the reflection coefficient $R$ will be concentrated in the exponent. In this way, the relationship of quantum entanglement with the reflection coefficient $R$ was found. Figure \ref{fig_5} shows the dependence of quantum entanglement for the Von-Neumann entropy $S_N$ and the Schmidt parameter $K$, depending on the reflection coefficient $R$.
For example, for $ s_1 = 1 $ and $ s_2 = 1 $, quantum entanglement will be in the form
\begin{eqnarray}
S_N=-(1-2R)^2\ln (1-2R)^2-4R(1-R)\ln \left( 2R(1-R)\right),
\nonumber\\
K=\frac{1}{1-8R(1-R)(1-3R(1-R))}.
\label{31}
\end{eqnarray}
For Eq. (\ref{31}) it is interesting to find the value $R$ at which there will be the maximum quantum entanglement and this value $R=1/2 (1\pm 1/\sqrt{3})$. At this value $R$, the quantum entanglement will be $S_N=\ln3$ and $K=3$. This is quite an interesting result, since at first glance it seems that the maximum entanglement should be at $R=1/2$. Fig. \ref{fig_5} shows that quantum entanglement strongly depends on the reflection coefficient $R$, but is always zero at $R=0,1$. The larger the quantum numbers $s_1, s_2$, the greater the quantum entanglement. You can see that there are two maxima quantum entanglement for different pairs $s_1, s_2$, except for the case when one of the quantum numbers is zero. No matter what measure of quantum entanglement we use, all dependencies are similar.

\begin{figure}[h]
\includegraphics[angle=0,width=0.9\textwidth, keepaspectratio]{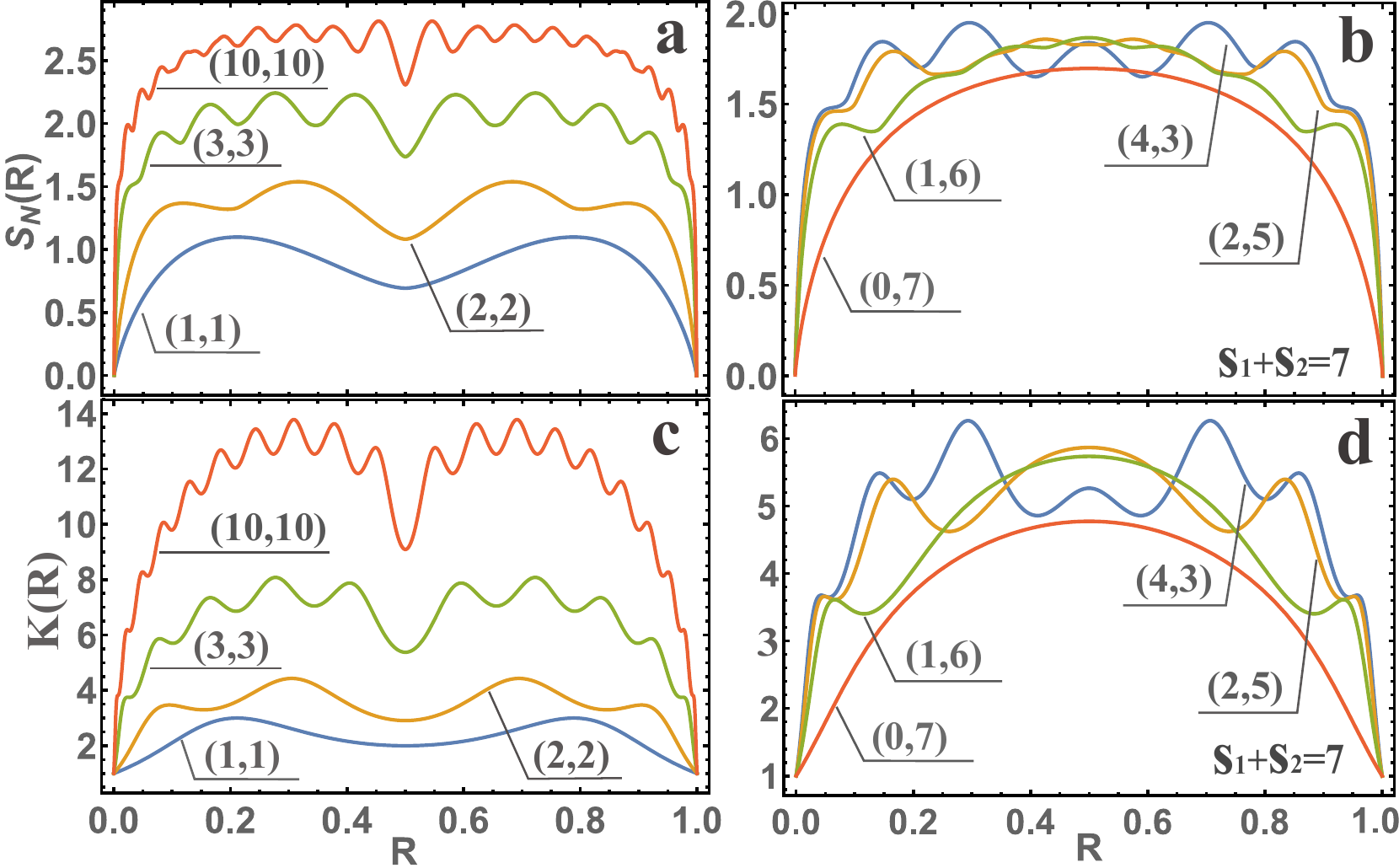}
\caption[fig_6]{The dependence of the von Neumann $ S_N $ entropy is shown in Fig. (a), as well as the Schmidt parameter in Fig. (b) for various values of $ R $ and $ s $.}
\label{fig_6}
\end{figure}

We also present quantum entanglement for various values of $s_1=s_2=s$ in Fig. \ref{fig_6}. The calculations are given for $R\in (0,1/2)$, since for symmetric values of $R\in (1/2,1)$, the results will be the same.
Figure \ref{fig_6} shows that the quantum entanglement at $s_1=s_2=s$ increases significantly with increasing $s$. For the Schmidt parameter $K$ , this dependence is close to linear.
Besides being able to obtain simple expressions for quantum entanglement, other physical characteristics can be calculated using equation (\ref{30}). For example, if one of the quantum numbers is zero, choose $s_2=0$, then in a very simple form one can find the average of the quantum numbers $k$ and $p$ (see Eq. (\ref{30})). The result is $\overline{k}=\sum_k k \lambda_k (R)=s_1(1-R)$ and $\overline{p}=\sum_p p \lambda_p (R)=s_1 R$.

One of the interesting applications of the results obtained is the possibility of obtaining the wave functions $ \Psi_{out}$  Eq. (\ref{29}) that have practical applications by specifying the coefficient $ R $. For example, we need a wave function $ \Psi (t) $ that defines the states of Holland-Burnett (HB)\cite{Holland_1993}. It is well known that this wave function is of great interest in various fields of physics, for example, in quantum metrology \cite{Polino_2020,Pezze_2018}. To do this, we need to select $ R = 1/2 $ and $ s_1 = s_2 = s $ (for even values $ s $). As a result, we get the wave function $ \Psi_{out}$ in the form
\begin{eqnarray}
\Psi =\sum^{s}_{n=0} e^{2 i n \phi}\frac{\sqrt{(2n)!(2s-2n)!}}{2^s n! (s-n)!} |2n,2s-2n \rangle .
\label{32}
\end{eqnarray}
It should be added that using the Eq. (\ref{32}) one can obtain an expression for quantum entanglement using the Schmidt parameter $ K $ in the form \cite{Makarov_PRE_2020}
\begin{eqnarray}
K=\frac{\pi (s!)^2}{\Gamma(s+1/2)^2 {_{4}}F_{3}(1/2,1/2,-s,-s;1,1/2-s,1/2-s;1)} ,
\label{33}
\end{eqnarray}
where $\Gamma(x)$ is the gamma function, ${_{4}}F_{3}(x_1,x_2,x_3,x_4;y_1,y_2,y_3;1)$ is the generalized hypergeometric function. It should be added that the Eq.(\ref{33}) has a fairly simple approximation $ K = s^{0.897} $. This clearly shows that quantum entanglement is unbounded from above, which was well known earlier (the more $ s $, the more the quantum entanglement).

If we consider the case when in the initial state $\Psi_{in}$ the number of photons in 1 mode $ s_1 $ and in the second one $ s_2 = 0 $. In this case the expression for quantum entanglement in the form of the Schmidt parameter will be
\begin{eqnarray}
K= \frac{1}{(1-R)^{2s_1} {_2F}_1\left(-s_1,-s_1;1;\left( \frac{R}{1-R} \right)^2\right)} ,
\label{34}
\end{eqnarray}
where ${_2F}_1(a,b;c;x)$ is Gaussian hypergeometric function. The general dependence of the Eq. (\ref {34}) can be seen from Fig.\ref{fig_5} (b, d). Also, analyzing the Eq.(\ref{34}), you can get that the maximum of this function at $ R = 1/2 $. With this value of $ R = 1/2 $, one can obtain a simpler expression for quantum entanglement 
\begin{eqnarray}
K_{max}= 2^{2s_1}\frac{(s_1!)^2}{(2 s_1)!}.
\label{35}
\end{eqnarray}
You can also find from the Eq. (\ref{35}) the parameter $ K $ for large values of the quantum number $ s_1 $, we get $ K_{max}(s_1 \gg1) \to \sqrt {\pi s_1} $. It can be seen in this case that the quantum entanglement is unbounded from above (the more $ s_1 $, the more the quantum entanglement).

It should be added that the cases of quantum entanglement given here using BS are only some cases, in reality there are many more.

\subsection{Quantum entanglement on a waveguide beam splitter}
Quantum entanglement on a frequency-dependent waveguide BS was recently considered in \cite{ Makarov_SR2_2021, Makarov_Enropy_2022}. In these papers, it was shown that quantum entanglement would be defined in the same way as in the ``Conventional'' waveguide BS, with the only difference that the Schmidt mode $\lambda_k$ must be replaced by $\lambda_k \to \Lambda_k $, where
\begin{eqnarray}
\Lambda_k= \int |\phi(\omega_1, \omega_2)|^2\lambda_k (R) d\omega_1 d \omega_2 ,~~ \lambda_k (R)=\left|c_{k,s_1+s_2-k}\right|^2.
\label{36}
\end{eqnarray}
We will assume that there is no quantum entanglement of photons at the BS input ports. In other words, we will consider the incoming photon states as Fock states, but photons are not monochromatic. In this case, as is well known, the photon wave function is factorizable, i.e. $\Psi_{in} =\int \phi_1 (\omega_1) |s_1 \rangle d \omega_1 \int\phi_2 (\omega_2) |s_2 \rangle d \omega_2 $, where $\phi(\omega_1,\omega_2)=\phi_1(\omega_1) \phi_2(\omega_2)$. 
To analyse quantum entanglement we will use von Neumann entropy $ S_N = - \sum_{k} \Lambda_k \ln \left (\Lambda_k \right) $. 
Next, let's choose $\phi_i (\omega_i) $ ($ i = 1,2 $) in the most commonly used form, this is the Gaussian distribution
\begin{eqnarray}
\phi_i(\omega_i)=\frac{1}{(2\pi)^{1/4}\sqrt{\sigma_i}}e^{-\frac{(\omega_i-\omega_{0i})^2}{4\sigma^2_i}},
\label{37}
\end{eqnarray}
where $\omega_{0i}$ is the mean frequency and $\sigma_i^2$ is the dispersion. Then we will assume that $\omega_{0i}/\sigma_i \gg 1$, which is applicable to most photon sources.

\begin{figure}[h!]
\center{\includegraphics[angle=0, width=0.93\textwidth, keepaspectratio]{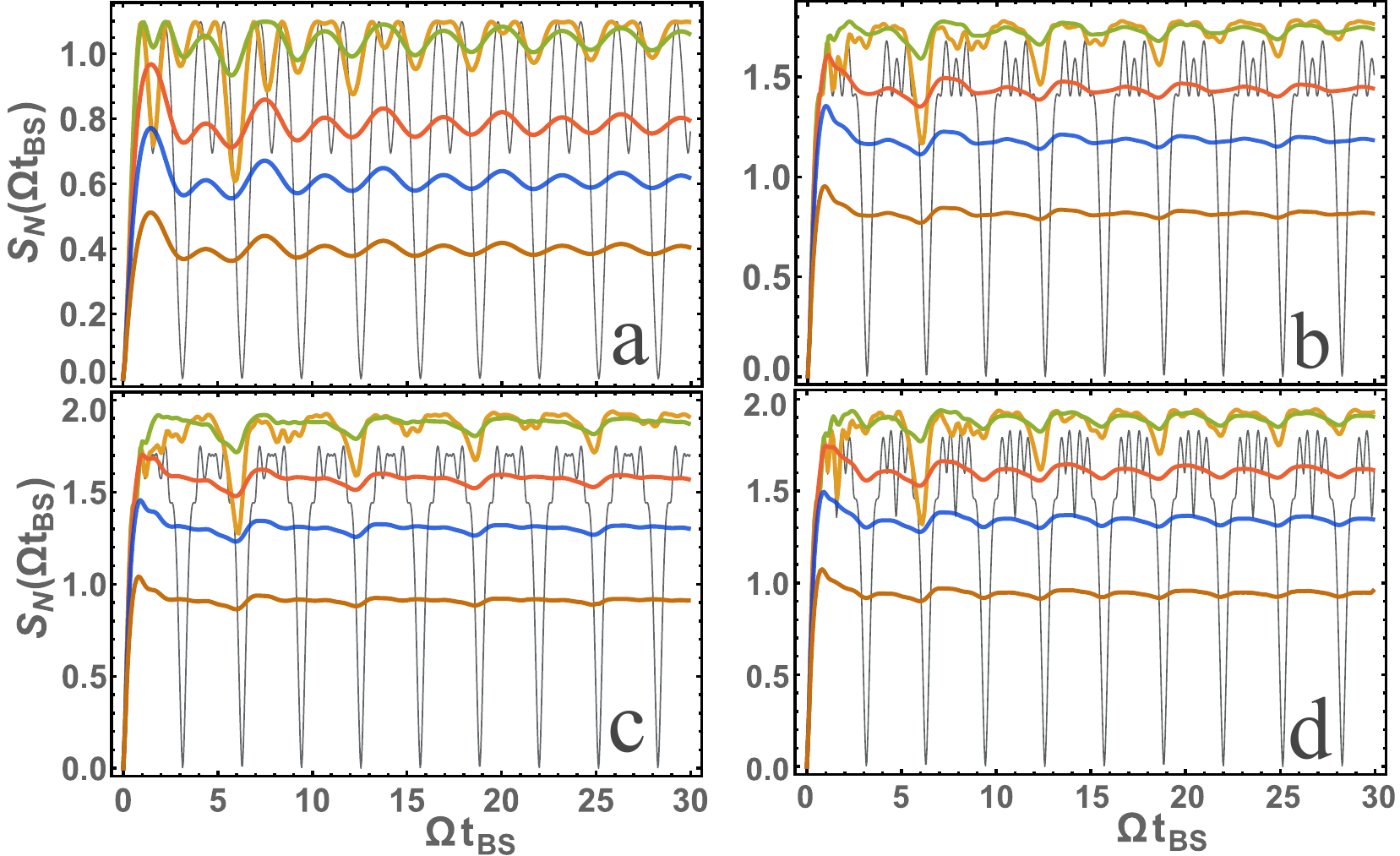}} 
\caption[fig_7]{The dependence of the von Neumann entropy $ S_N $ on the parameter $ \Omega t_ {BS} $ is presented. In figures (a, b, c, d) the value of $ S_N $ is presented for $ | 1,1 \rangle, | 2,3 \rangle, | 4,2 \rangle, | 3,3 \rangle $, respectively (where $ | s_1, s_2 \rangle $ are input states to 1 and 2 ports of the BS, respectively). All figures show the results for $ \sigma / \Omega = 10$ (brown), $\sigma / \Omega = 5$ (blue),$\sigma / \Omega = 3$ (red), $\sigma / \Omega = 1$ (green), $\sigma / \Omega = 1/3$ (orange) from bottom to top, respectively. The thin curve is made at $\sigma / \Omega = 0$ (black).}
\label{fig_7}
\end{figure}

Fig. \ref{fig_7} let us represent the dependence of the von Neumann entropy $ S_N $ depending on the dimensionless parameter $ \Omega t_{BS} $ for identical photons, i.e. for $ \sigma_1 = \sigma_2 = \sigma $ and $ \omega_{01} = \omega_{02} = \omega_{0} $. 
We add that the results obtained for monochromatic photons, i.e. for $ \sigma \to 0 $ (more precisely, for $ \sigma / \Omega \ll 1 $ such that $ \sigma t_{BS} \ll 1 $) are in agreement with the results of quantum entanglement ``conventional'' BS calculations, see for example \cite{Kim_2002, Makarov_PRE_2020}. Indeed, if we use the results for ``conventional'' BS with reflection coefficient $ R = \sin^2(\kappa z ) $ and, for example, \cite{Kim_2002} to calculate quantum entanglement (or the above method), we can get results in Fig. \ref{fig_7} for $ \sigma / \Omega = 0 $ (thin lines). Thus our result is a more general one applicable to non-monochromatic photons.
One can see that there is a big difference between quantum entanglement of monochromatic and non-monochromatic photons. Moreover, in the case of non-monochromatic photons, when $ \sigma / \Omega \sim 1 $, the quantum entanglement is larger. One can also see that in the case of non-monochromatic photons, when $ \sigma / \Omega \gtrsim 1 $ and for relatively large $ \Omega t_{BS} $, the quantum entanglement becomes a constant value. In the case of monochromatic photons quantum entanglement is a periodic function. 

\begin{figure}[]
\center{\includegraphics[angle=0, width=0.9\textwidth, keepaspectratio]{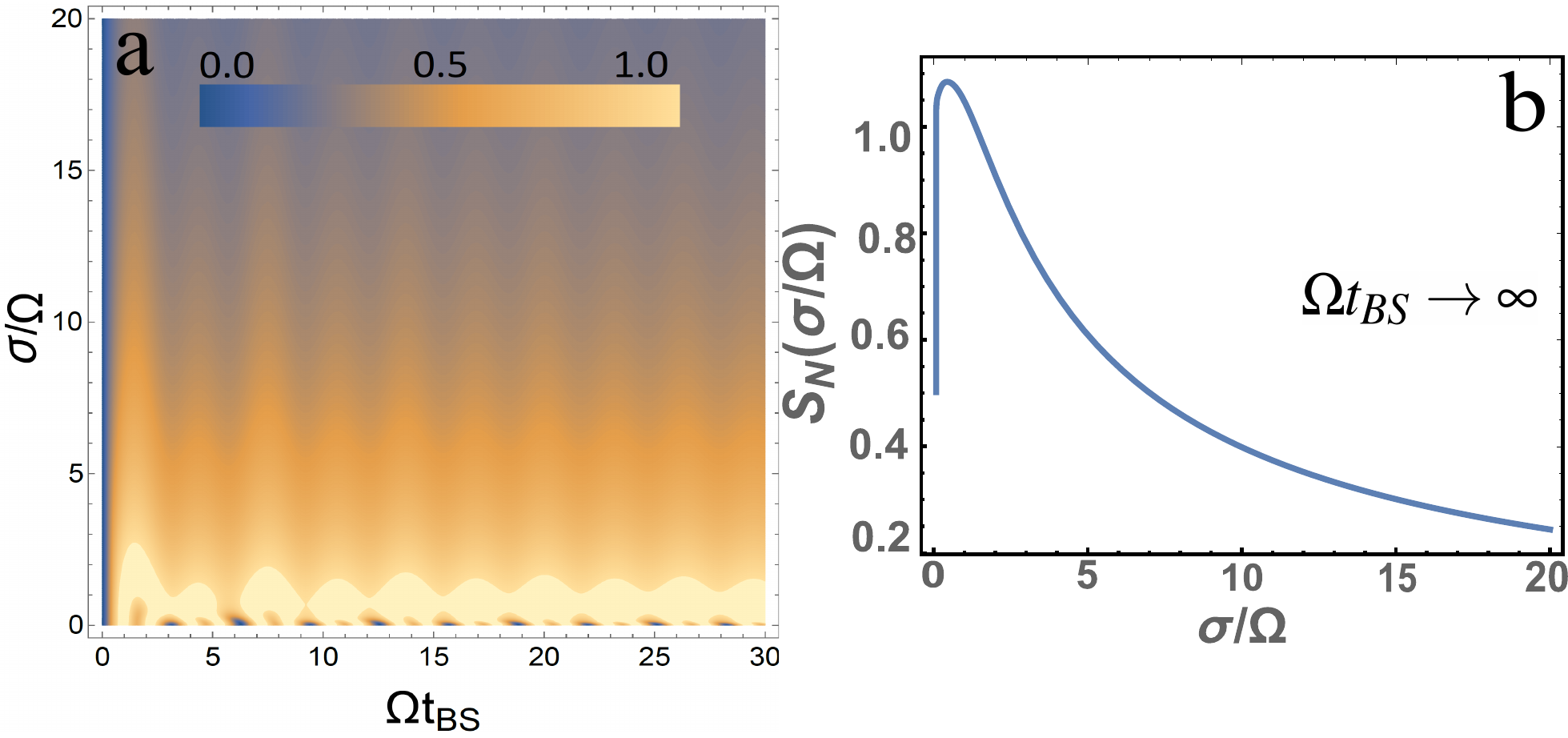}} 
\caption[fig_8]{(a): A contour plot of the von Neumann entropy $ S_N $ versus two parameters $ \sigma / \Omega $ and $ \Omega t_ {BS} $ for the input state $ | 1, 1 \rangle $ is presented. (b): von Neumann enetropy $ S_N $ is presented as a function of $ \sigma / \Omega $ in the limiting case $ \Omega t_{BS} \to \infty $ for the input state $ | 1, 1 \rangle $, see \cite{Makarov_SR2_2021}}
\label{fig_8}
\end{figure}

\begin{figure}[!]
\center{\includegraphics[angle=0, width=0.95\textwidth, keepaspectratio]{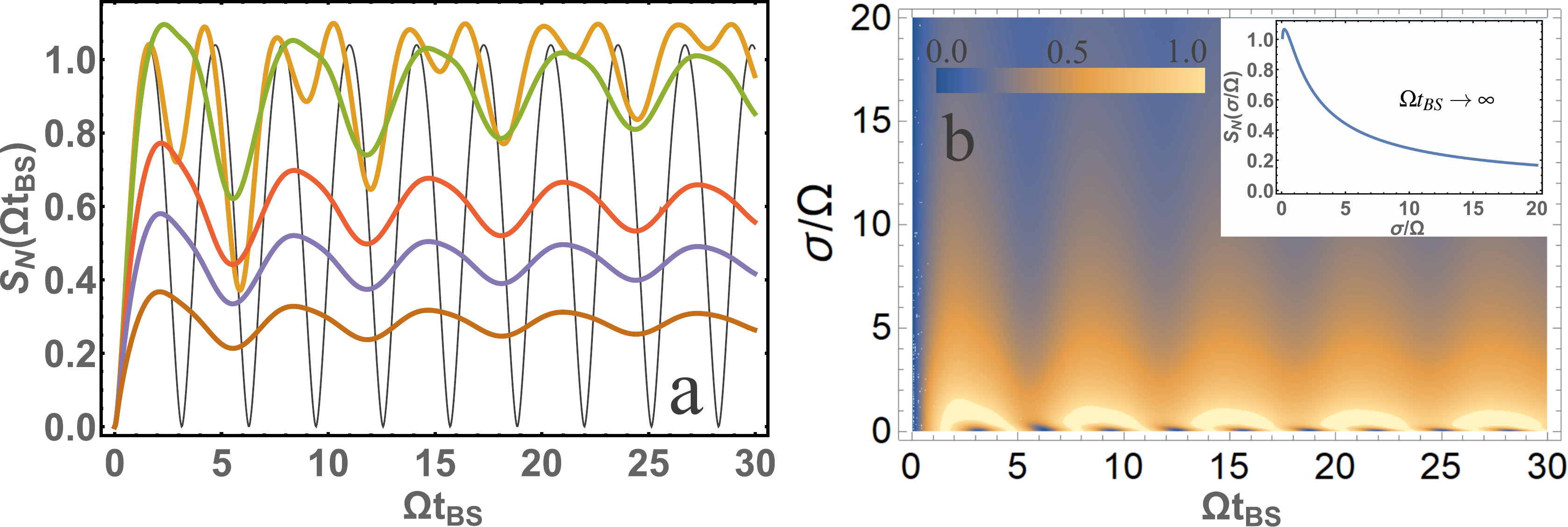}} 
\caption[fig_9]{(a): The dependence of Von Neumann entropy $S_{N}$ on the parameter $\Omega t_{BS} $ is presented for $\sigma/\Omega=10$ (brown); $\sigma/\Omega=5$ (blue); $\sigma/\Omega=3$ (red); $\sigma/\Omega=1$ (orange); $\sigma/\Omega=1/3$ (green); $\sigma/\Omega=0$ (black). (b): A contour plot of Von Neumann entropy $S_{N}$ from two system parameters $\Omega t_{BS}$ and $\sigma/\Omega$ is presented. The inset is presented for $S_{N}$ at $\Omega t_{BS} \to \infty$ depending on the parameter $\sigma/\Omega$. Input photons are in the  state $|0\rangle,|2\rangle$, see \cite{Makarov_Enropy_2022}.}
\label{fig_9}
\end{figure}

As an example, let us consider in more detail the case of input photons in states $ | 1, 1 \rangle $. This case is interesting because it realizes the case of the Hong-Ou-Mandel (HOM)\cite{HOM_1987, Makarov_OL_2020} effect. To analyze this case, it is convenient to present a contour plot for von Neumann entropy as a function of two parameters $ \sigma / \Omega $ and $ \Omega t_{BS} $, see Fig. \ref{fig_8}(a). Since at large $ \Omega t_{BS} $ (of course, the condition $ \sigma t_ {BS} \gg 1 $ must also be satisfied) the quantum entanglement tends to a constant value, in Fig. \ref{fig_8}(b) presents this constant value depending on the parameter $ \sigma / \Omega $. It is also possible to find an analytic dependence for quantum entanglement for the case $|1,1\rangle$ as
\begin{eqnarray}
 S_N=\ln \frac{2(1-{\rm J})^{{\rm J}-1}}{(2{\rm J})^{{\rm J}}} ,~~     {\rm J}=1+\frac{3}{8}\left( \frac{\Omega}{\sigma}\right)^2-\frac{\sqrt{\pi}}{16} \left( \frac{\Omega}{\sigma}\right)^3  \left\{3+10 \left( \frac{\sigma}{\Omega}\right)^2 \right\}{\rm erf}\left( \frac{\Omega}{2\sigma}\right){e}^{\left( \frac{\Omega}{2\sigma}\right)^2} ,
\label{38}
\end{eqnarray}
where $\texttt{erf}$ is an error function. Should be added that, to get the Eq. (\ref{8}) sine and cosine terms (rapidly oscillating terms at $ \Omega t_{BS} \to \infty $) must be ignored when integrating over frequencies.
It can be seen from the graphs obtained that the quantum entanglement has a maximum, which will be at $ \sigma / \Omega = 0.44467 $. It is also seen that the quantum entanglement is large at $ \sigma / \Omega \sim 1 $, and as $ \sigma / \Omega $ increases, it tends to zero. As an example, let us give another calculation of quantum entanglement from the initial states $|0,2\rangle$ at the same remaining parameters, see Fig. \ref{fig_9}.
Fig. \ref{fig_9} shows that the Von Neumann entropy at $\sigma/\Omega \gtrsim 1$ is quite different from the case of monochromatic photons $\sigma/\Omega =0$. Fig. \ref{fig_9} (b) also shows that the large value of entropy is at $\sigma/\Omega \sim 1$. The largest entropy value at $\Omega t_{BS} \to \infty$ would be $S_N=1.092$ at $\sigma/\Omega=0.24$.

In summary, it can be concluded that the waveguide BS can be a good source of quantum entangled photons. And the entanglement can be easily adjusted by changing the $\Omega$ parameter, e.g. by separating or bringing the waveguides closer together. It should be noted that a waveguide BS can be a source of large quantum entangled photons. Indeed, such a source generates almost maximum possible quantum entanglement at $\sigma / \Omega \sim 1 $ and $\Omega t_ {BS}> 1 $. Recall that the maximum quantum entanglement for von Neumann entropy in our case $ S_N = \ln (1 + s_1 + s_2) $ \cite{Kim_2002, Phoenix_1988}. If we consider constants $ R $ and $ T $ (monochromatic photons) the quantum entanglement is a periodic function of $ \Omega t_ {BS} $, and for large $ \Omega t_{BS} \gg 1 $ a rapidly oscillating relation, which is bad for use in quantum technologies.
\section{Photon statistics on the beam splitter}
The statistical properties at the BS output ports are determined by the probability $P_k$ to detect $k$ photons on port 1 and the probability $P_p$ to detect $p$ photons on port 2. The statistical properties at the BS output ports have been studied in many works \cite{Campos_1989,Makarov_PRE_2020,Makarov_SR2_2021, Kim_2002,Ou_2007}. Initial states may depend on which photons are fed to the BS input ports. These states may differ not only in the distribution of photons in each mode, but also in the quantum entanglement of the electromagnetic modes of the incoming photons. We will limit ourselves here to considering the most important example - these are the incoming Fock states, since other cases can be represented based on them. 
\subsection{Photon statistics on a ``conventional'' beam splitter}
\begin{figure}[h!]
\center{\includegraphics[angle=0, width=0.8\textwidth, keepaspectratio]{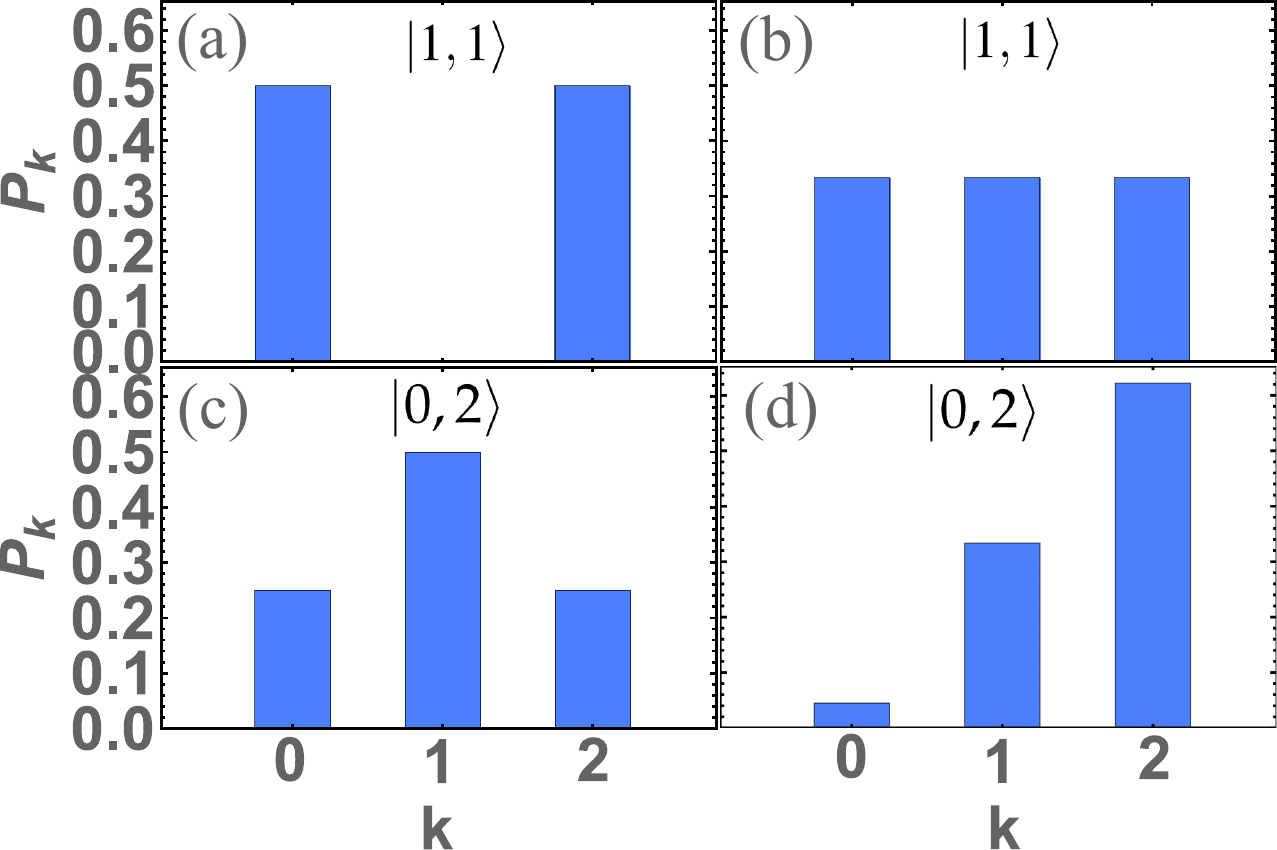}} 
\caption[fig_10]{The calculation of probability $P_k$ for the case $|1,1\rangle$ in Figures (a) at $R=1/2$ and (b) at $R=1/2(1+1/\sqrt{3})$ and for the case $|0,2\rangle$ in Figures (c) at $R=1/2$ and (d) at $R=1/2(1+1/\sqrt{3})$ are presented.}
\label{fig_10}
\end{figure}

In the case of ``conventional'' BS, the statistical properties at the BS output ports have been well studied in \cite{Campos_1989}.  The simplest expressions for calculating statistical properties have been derived in \cite{Makarov_PRE_2020}, which are defined by Eq. (\ref{30}), where the probability $P_k=\lambda_k$ and the probability $P_p=\lambda_{s_1+s_2-k}$, since the number of photons is conserved, i.e. $s_1+s_2=k+p$. As an example, let's present the calculations for the case $|1,1\rangle$ in Fig. \ref{fig_10} (a) and (b) and for the case $|0,2\rangle$ in Fig. \ref{fig_10} (c) and (d) for the two reflection coefficient values $R=1/2,1/2(1+1/\sqrt{3})$. The choice of these values  $R$ is determined by the fact that for $|1,1\rangle$ and $R=1/2$ the HOM effect occurs, and at $R=1/2(1+1/\sqrt{3})$ the quantum entanglement value is maximum (see Eq.(\ref{31})). 
It can be seen that the HOM effect does not appear at the maximum entanglement of photons, but at the maximum entanglement the distribution is equally probable. It should also be added that calculating the probabilities for any initial states is not difficult using Eq. \ref{30}).
\subsection{Photon statistics on a waveguide beam splitter}
\begin{figure}[h!]
\center{\includegraphics[angle=0, width=0.8\textwidth, keepaspectratio]{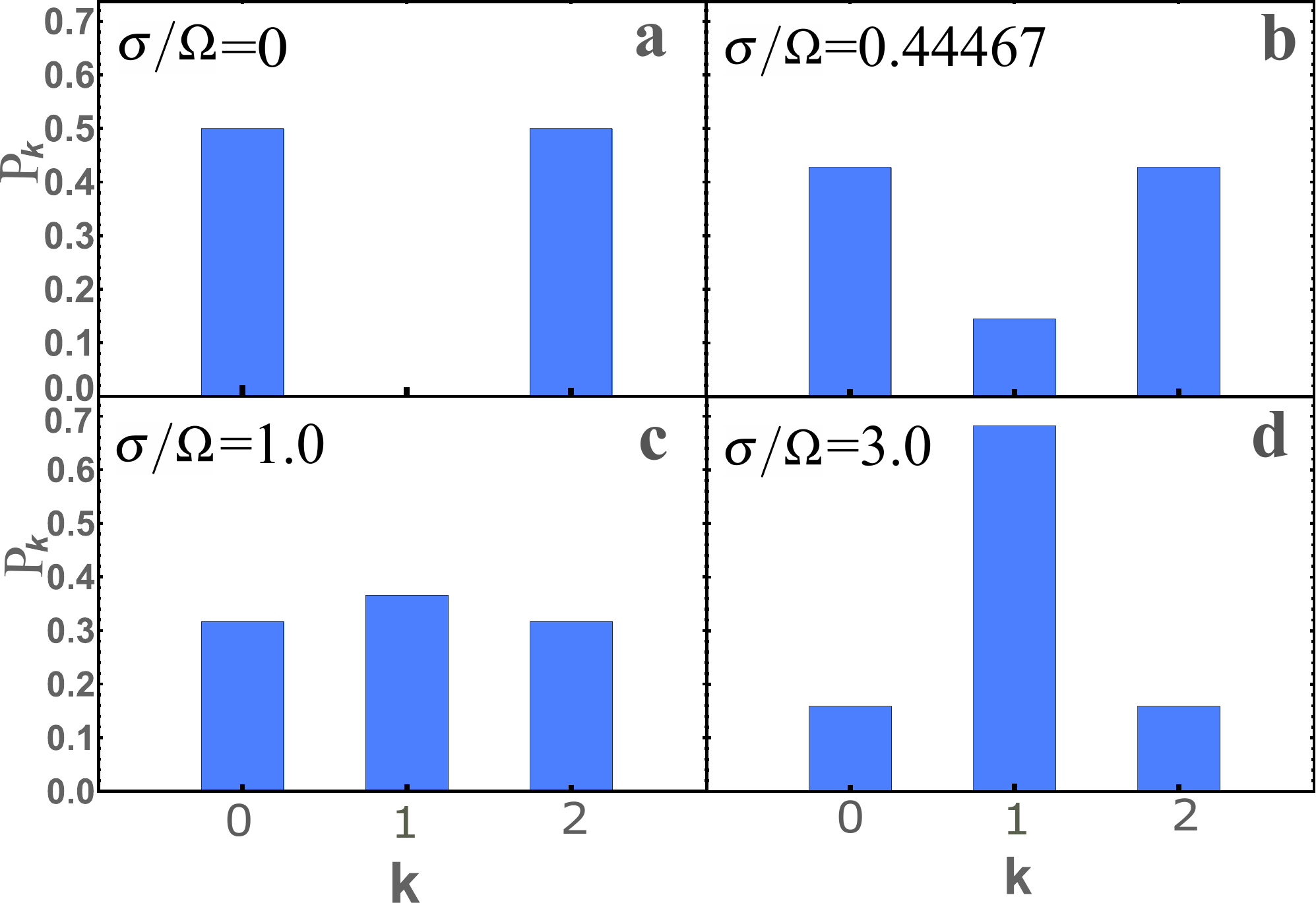}} 
\caption[fig_11]{A histogram of the dependence of the probability $ P_k $ of detecting $ k $ and $ p = s_1 + s_2-k $ (we consider the case $ | 1,1 \rangle $, where $ s_1 = s_2 = 1 $) photons at the output of the first and second ports, respectively, for different values of $ \sigma / \Omega $.}
\label{fig_11}
\end{figure}

\begin{figure}[h!]
\center{\includegraphics[angle=0, width=0.98\textwidth, keepaspectratio]{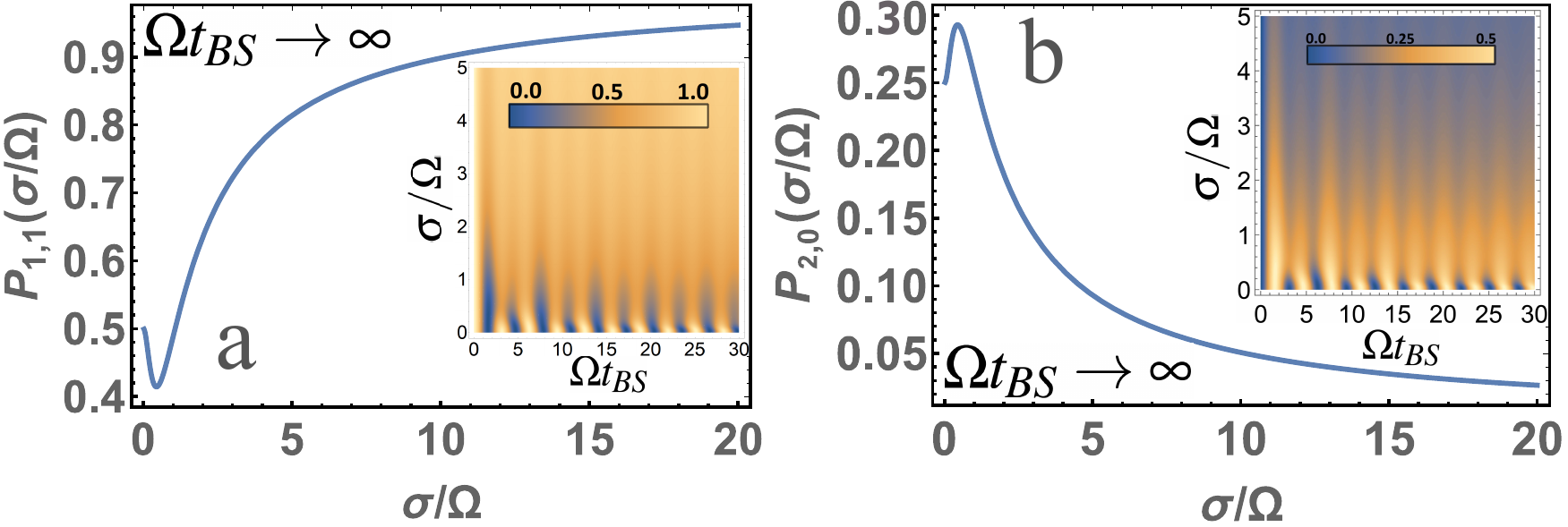}} 
\caption[fig_5]{Figure (a) shows the probability $ P_{1,1} $, Figure (b) shows the probability $ P_{2,0} = P_ {0,2} $ for $ \Omega t_{BS} \to \infty $ depending on the $ \sigma / \Omega $ parameter.
Also in the insets Figure (a),(b) are contour plots for the probability $ P_{1,1} $ and $ P_ {2,0} = P_ {0,2} $ are presented, respectively, depending on two parameters of the studied system $ \sigma / \Omega $ and $ \Omega t_{BS} $. }
\label{fig_12}
\end{figure}

Let us consider the photon statistics at the BS output ports. Expressions to calculate the statistical properties were obtained in \cite{Makarov_SR2_2021}, which are defined by Eqs. (\ref{30}) and (\ref{36}), where probability $P_k=\Lambda_k$ and probability $P_p=\Lambda_{s_1+s_2-k}$.
It should be added that when considering monochromatic photons the coefficients $ R $ and $\phi $ would be constant, then following Eq. (\ref{36}) $ \Lambda_k = \lambda_k $.  Thus, our consideration is general and in the particular case coincides with ``conventional'' BS. Let us compare the results in the case of monochromatic and non-monochromatic photons. 
To do this let us choose BS length $ z = v t_{BS} $ and frequency $ \Omega = 2 \kappa v $ (see below, after Eq. (\ref{28})) so that the reflection coefficient for monochromatic and identical photons is $ R = \sin^2 (\Omega t_{BS}/2) = 1/2 $. In this case we will choose $ \Omega t_{BS} = 5 \pi / 2 $. Next we will consider the same ($ \sigma_1 = \sigma_2 = \sigma $ and $ \omega_{01} = \omega_{02} = \omega_{0} $) but not monochromatic photons, see Eq. (\ref{37}). We present for comparison the calculation results in the case of identical monochromatic photons for $R=1/2$ (at $ \Omega t_{BS} = 5 \pi / 2 $) and in the case of identical but not monochromatic photons (at $ \Omega t_{BS} = 5 \pi / 2 $), see Fig. \ref{fig_11}. From Fig. \ref{fig_11}(a) shows that for states $ | 1,1 \rangle $ when $ \sigma / \Omega = 0 $ the HOM effect \cite{HOM_1987} is realized, see Fig. \ref{fig_10} (a). This means that only photon pairs are detected at the first or second detector (in the figure it is for $ k = 0 $ and $ k = 2 $) with probability $ 1/2 $. With increasing $ \sigma / \Omega $ the HOM effect disappears and the photon statistics change dramatically. It is quite interesting to look at photon statistics at maximum quantum entanglement, see Fig. \ref{fig_11}(b). These statistics are different from the statistics of the HOM effect of Fig. \ref{fig_11}(a) and from the statistics in the case of monochromatic photons at maximum quantum entanglement, see Fig. \ref{fig_10}(b). At large $\sigma / \Omega \gg 1 $, the probability $\Lambda_k $ will tend to one for $ k = 1 $. This means that photons will never arrive at detectors in pairs. A similar analysis can quite easily be done for any input states $ | s_1, s_2 \rangle $. The commonality for all cases will be a coincidence with the statistics for monochromatic photons at $\sigma /\Omega = 0 $. Also, the general probability behaviour for any states $ | s_1, s_2 \rangle $ will be at $ \sigma / \Omega \gg 1 $. This is easily explained because at $ \Omega \to 0 $ (same as $ \sigma / \Omega \gg 1 $), the coupling between the waveguides in the BS weakens, which means that photons propagate along their waveguides.
Also consider the limiting case described above for quantum entanglement, this is the case for large $ \Omega t_ {BS} $ (more precisely, the condition $ \sigma t_{BS} \gg 1 $ must also be satisfied) for the input state $ | 1,1 \rangle $. In this case, it is easy to obtain the probability of detecting photons on both detectors $ P_{1,1} = {\rm J} $, where $ {\rm J} $ is represented in Eq. (\ref {38}). The probability of detecting pairs of photons on 1 or 2 detectors will then be $ P_ {2,0} = P_ {0,2} = 1/2 (1-P_ {1,1}) $. The results are presented in Fig.\ref{fig_12}.
It is interesting enough to note that the minimum of the function $ min \{P_{1,1} \} $ or the maximum of $ max \{ P_{2,0} = P_{0,2} \} $ for $ \sigma / \Omega = 0.44029 $ practically coincides with the maximum for quantum entanglement at $ \sigma / \Omega = 0.44467 $, see Fig.\ref{fig_8}(b). This means that the maximum of quantum entanglement is realized when photons at the first and second detectors can be registered with a minimum probability. Or, which is the same, when pairs of photons can be recorded on detectors with maximum probability. Also in the insets Fig.\ref{fig_12}(a),(b) the probabilities $ P_{1,1} $ and $ P_ {2,0} = P_ {0,2} $ are presented, respectively, depending on two parameters of the studied system $ \sigma / \Omega $ and $ \Omega t_{BS} $. It can be seen that taking into account the non-monochromaticity of photons significantly changes the probabilities, in comparison with monochromatic ones.
It should be added that such an analysis, for quantum entanglement and photon statistics, is fairly easy to carry out for any photons input states $ | s_1, s_2 \rangle $, as well as various BS parameters $ \Omega, t_{BS} $ and the values of non-monochromaticity of photons $ \sigma_1, \sigma_2 $.

\section{Hong-Ou-Mandel effect}
The HOM effect was first experimentally demonstrated by Hong et al. in 1987 \cite{HOM_1987}. The effect is thought to occur when two identical single-photon waves hit the BS 1:1 (with reflection coefficients $ R $ and transmittance $ T $ close to 1/2), one at each input port. When the photons are identical, they will cancel each other out. HOM interference appears in many cases, both in fundamental studies of quantum mechanics and in practical implementations of quantum technologies \cite{Gisin_2002,Sangouard_2011}.  
The theoretical explanation of the HOM effect, based on constant coefficients $ R $ and $ T $ and bosonic photon statistics, is quite simple \cite{Mandel_1995,Agarwal_2013}. In this interpretation we are not interested in what happens to the incident photons in the BS. For this, we consider a lossless BS with constant coefficients $ R $ and $ T $ (i.e.  ``conventional'' BS) and the BS is the source of two other photons obeying bosonic statistics.  In this case, the wave function at the output ports $\Psi_{out}={{\hat b_1}^{\dagger}} {{\hat b_2}^{\dagger}}|0\rangle$, where ${\hat b_1},{\hat b_2}$ is defined through the BS matrix as Eq. (\ref{3}). It's easy to see that ${{\hat b_1}^{\dagger}} {{\hat b_2}^{\dagger}}=(T-R){{\hat a_1}^{\dagger}}{{\hat a_2}^{\dagger}}+\sqrt{RT}\left(e^{-i\phi}\left({\hat a_2}^{\dagger}\right)^2-e^{i\phi}\left({\hat a_1}^{\dagger}\right)^2 \right)$. Choosing the coefficients $R=T=1/2$ we get ${{\hat b_1}^{\dagger}} {{\hat b_2}^{\dagger}}=1/2\left(e^{-i\phi}\left({\hat a_2}^{\dagger}\right)^2-e^{i\phi}\left({\hat a_1}^{\dagger}\right)^2 \right)$. It means that photons get to detectors only in pairs, i.e. the probability will be $P=1/2$ for each of detectors, that contradicts to the classical representation of separation of two beams of light with coefficients $R=T=1/2$. In the classical notion at $R=T=1/2$ there can be 4 variants: \\
1. first and second photons fall on detectors 1 and 2 accordingly \\
2. first and second photons fall on detectors 2 and 1 accordingly \\
3. first and second photons fall on detector 1 \\
4. first and second photons fall on detector 2. \\
A total of 4 equally probable events, which obviously gives a probability of $P=1/4$ for each of the 4 choices. Such phenomenon is called the HOM effect (or HOM interference) and it is a good method to check quantum properties not only of photons, but also of other particles. In other words, in the HOM effect, the probability of photons hitting the first and second detector $P_{1,2}=(R-T)^2$, with equal $R$ and $T$ will be zero. If we consider non-monochromatic photons but identical (frequency dependent BS), it is easy to see that $P_{1,2}=\int |\phi(\omega_1, \omega_2)|^2\left(R(\omega_1,\omega_2)-T(\omega_1,\omega_2)\right)^2 d \omega_1 d\omega_2=\overline{(R-T)^2}$. Choosing $\bar{R}=\bar{T}=1/2$ we can get $ P_{1,2} = 4 (\overline {T^2} - \overline{T}^2) = 4 (\overline{R^2} - \overline{R}^2) $, i.e. there is a fluctuation of the reflectance $R$ and transmittance $T$, which was not considered before. These issues will be discussed in detail here. 
\subsection{Hong-Ou-Mandel effect on a ``conventional'' beam splitter}
A schematic of the HOM interference experiment can be shown in Fig.\ref{fig_4}, with the difference that we consider ``conventional'' BS. The detector coincidence rate will drop to zero when identical input photons completely overlap in time. This is called the Hong-Ou-Mandel dip, or HOM dip. Let us consider the general case not restricted to identical and monochromatic photons and to simultaneous registration by photon detectors. In this case we have to consider the output state as (see Eq.(\ref{6})) $|\Psi_{out}\rangle=\int \phi(\omega_1, \omega_2){{\hat b_1}^{\dagger}} {{\hat b_2}^{\dagger}}|0\rangle d\omega_1 d\omega_2$, where ${\hat b_1}, {\hat b_2}$ is defined by Eq.(\ref{27}). It should be added that getting Eq. (\ref{27}) to find ${\hat b_1}(t_1), {\hat b_2}(t_2)$ quite easily if we consider ${\hat b_k}(t_k)=e^{i {\hat H_0}t_k}{\hat b_k}(0)e^{-i {\hat H_0}t_k}$, where ${\hat b_k}(0)={\hat b_k}$ from Eq. (\ref{3}), where $k=(1,2)$. We will assume the coefficients $R$ and $\phi$ to be constant, i.e. consider ``conventional'' BS.  Consider the probability $ P_ {1,2} $ of co-detection of photons at 1 and 2 detectors (correlation between the two detectors). If our coincidence gate window takes counts for time $ T_D $, then the frequency of coincidence between detectors 1 and 2 is proportional (see, e.g., \cite{HOM_1987, Fearn_1989, Aephraim_1992}).
\begin{eqnarray}
P_{1,2}\propto \int^{T_{D}/2}_{-T_{D}/2}\int^{T_{D}/2}_{-T_{D}/2}\langle {\hat b}^{\dagger}_1(t_1){\hat b}^{\dagger}_2(t_2){\hat b}_1(t_1){\hat b}_2(t_2) \rangle dt_1 dt_2.
\label{39}
\end{eqnarray}
Consider the case where the reaction time $ \tau_D $ (time resolution) of detectors $ D_1 $ and $ D_2 $ in the experiment is many times slower than other time scales of the problem (reaction time is large), then in this case $ T_D \to \infty $. It should be added that the theory presented below is not difficult to generalize to the case when the detector reaction times are short, which is currently realized experimentally (eg \cite{Legero_2004, Lyons_2018}). The calculation of the correlation function (\ref{39}) can be found in various papers depending on the JSA function $\phi(\omega_1, \omega_2)$, eg \cite{HOM_1987,Fearn_1989,Aephraim_1992,Wang_2006,Legero_2004,Branczyk_2017}. As a result, the correlation function in general form will be at $R=T=1/2$ 
\begin{eqnarray}
P_{1,2}=\frac{1}{2}\left(1- {\rm Re}\biggl\lbrace \int \phi(\omega_1,\omega_2)\phi^*(\omega_2,\omega_1)e^{-i(\omega_2-\omega_1)(\delta \tau +\tau)}\biggl\rbrace \right) d \omega_1 d\omega_2  ,
\label{40}
\end{eqnarray}
where $ P_{1,2} $ is normalised so that at $R=0$ the probability is $ P_{1,2} = 1 $ (without BS the probability of detection is $ 100 \% $), which corresponds to the standard normalisation in HOM theory. In Eq. (\ref{40}) is usually replaced by $\Delta \tau =\delta \tau+\tau$ corresponding to the time between detection of 1 and 2 photons. It can be seen that at $\Delta \tau=0$ the correlation function $P_{1,2}=0$, only when the two photons are identical, i.e. at $\phi(\omega_1,\omega_2)=\phi(\omega_2,\omega_1)$. Thus, the HOM effect will be observed at photon identity and at $\Delta \tau=0$. 

Here are examples of the calculated correlation functions $P_{1,2}$. Here we give the result obtained in the very first paper where this effect was demonstrated \cite{HOM_1987} 
\begin{eqnarray}
P_{1,2}=\frac{1}{2}\left( 1-e^{-(\delta \omega \Delta \tau)^2} \right),
\label{41}
\end{eqnarray}
where $\delta \omega$ is the bandwidth at Gaussian $\phi(\omega_0/2+\omega,\omega_0/2+\omega)$. It should be added that sources of single photons are often used in experiments, but their initial state cannot always be determined by the Fock states. In other words, photons incident on the BS can be quantum entangled. One of the most commonly used sources of such photons is spontaneous parametric down-conversion (SPDC). In the case of such photons, the spectral function $\phi(\omega_1,\omega_2)$ can have a complicated form. Quite a lot of such cases are considered in \cite{Wang_2006}.

Here we have considered photons to demonstrate the HOM effect. This effect has a deeper meaning than demonstrated here. First of all, this is due to the fact that the HOM effect can be observed not only on photons, but also on other particles of both bosonic and fermionic nature \cite{Liang_2005, Toyoda_2015,Aspect_2019}.

\subsection{Hong-Ou-Mandel effect on a waveguide beam splitter}
While the HOM effect theory for ``conventional'' BS is well known and described in various literature, the HOM effect theory for frequency-dependent waveguide BS has only recently appeared \cite{Makarov_OL_2020, Makarov_SR_2020}. This theory will be presented here in more detail.
Now the scheme of the HOM experiment will look like Fig. \ref{fig_4}. The calculation of the correlation function is carried out in the same way as it is done in the case of ``conventional'' BS, with the only difference that we consider the coefficients $R,T$ and $\phi$ to be frequency dependent and do not fix their values. As a result, we get
\begin{eqnarray}
P_{1,2}=\int^{\infty}_{-\infty}\int^{\infty}_{-\infty} \left|\xi_1(t_1,t_2,\tau)-\xi_2(t_1,t_2,\tau,\delta \tau) \right|^2  dt_1 dt_2~,
\nonumber\\
\xi_1(t_1,t_2,\tau)=\frac{1}{2\pi}\int \phi(\omega_1,\omega_2)T(\omega_1,\omega_2)e^{-i\omega_1(t_1-\tau)}e^{-i\omega_2 t_2} d\omega_1 d \omega_2~,
\nonumber\\
\xi_2(t_1,t_2,\tau,\delta \tau)=\frac{1}{2\pi}\int \phi(\omega_1,\omega_2) R(\omega_1,\omega_2) e^{-i\omega_2(t_1+\delta \tau)}e^{-i\omega_1(t_2-\delta \tau -\tau)} d\omega_1 d \omega_2 ~,
\label{42}
\end{eqnarray}
where $ P_{1,2} $ is normalized so that with $ t_{BS} = 0 $ the probability is $ P_{1,2} = 1 $. We then obtain
\begin{eqnarray}
P_{1,2}=\int \Biggr(  |\phi(\omega_1,\omega_2)|^2 \left( T^2(\omega_1,\omega_2)+R^2(\omega_1,\omega_2) \right)-
\nonumber\\
2{\rm Re}\biggr\lbrace \phi(\omega_1,\omega_2)
\phi^*(\omega_2,\omega_1) T(\omega_1,\omega_2)R(\omega_2,\omega_1)
e^{-i(\omega_2-\omega_1)\Delta \tau}\biggl\rbrace \Biggl) d \omega_1 d \omega_2  .
\label{43}
\end{eqnarray}
It should be added that if $ T $ and $ R $ are assumed to be independent of frequencies and $ T = R = 1/2 $, then Eq. (\ref{43}) corresponds to the well-known equation, eg \cite{Grice_1997,Erdmann_2000,Barbieri_2017}, see above. 

We next consider the case of identical photons at $ \Delta \tau = 0 $, in this case $ \phi (\omega_1, \omega_2) = \phi(\omega_2, \omega_1) $ and $ R(\omega_1, \omega_2) = R (\omega_2, \omega_1) $, and the quantity
\begin{eqnarray}
P_{1,2}(\Delta \tau=0)=\overline{(T-R)^2} =
\int   |\phi(\omega_1,\omega_2)|^2 \left( T(\omega_1,\omega_2)-R(\omega_1,\omega_2) \right)^2  d \omega_1 d \omega_2 .
\label{44}
\end{eqnarray}
If in (\ref{44}) we choose $ \overline{T} = \overline{R} = 1/2 $, then we get $ P_{1,2} = 4 (\overline {T^2} - \overline{T}^2) = 4 (\overline{R^2} - \overline{R}^2) $. In other words, there is a mean-square fluctuation of the coefficients of transmission $ T $ and reflection $ R $, which leads to a nonzero value of $ P_ {1,2} $ in the case of identical photons. This conclusion is fundamental in the theory of HOM interference and was not previously obtained. Also, from the previously obtained Eqs. (\ref{43}) and (\ref{44}) it follows that the $ P_{1,2}(\Delta \tau \gg \tau_c)=2\overline{T^2}=2\overline{R^2}$ ($ \tau_c $ is the coherence time).

\begin{figure}[h!]
\center{\includegraphics[angle=0, width=0.9\textwidth, keepaspectratio]{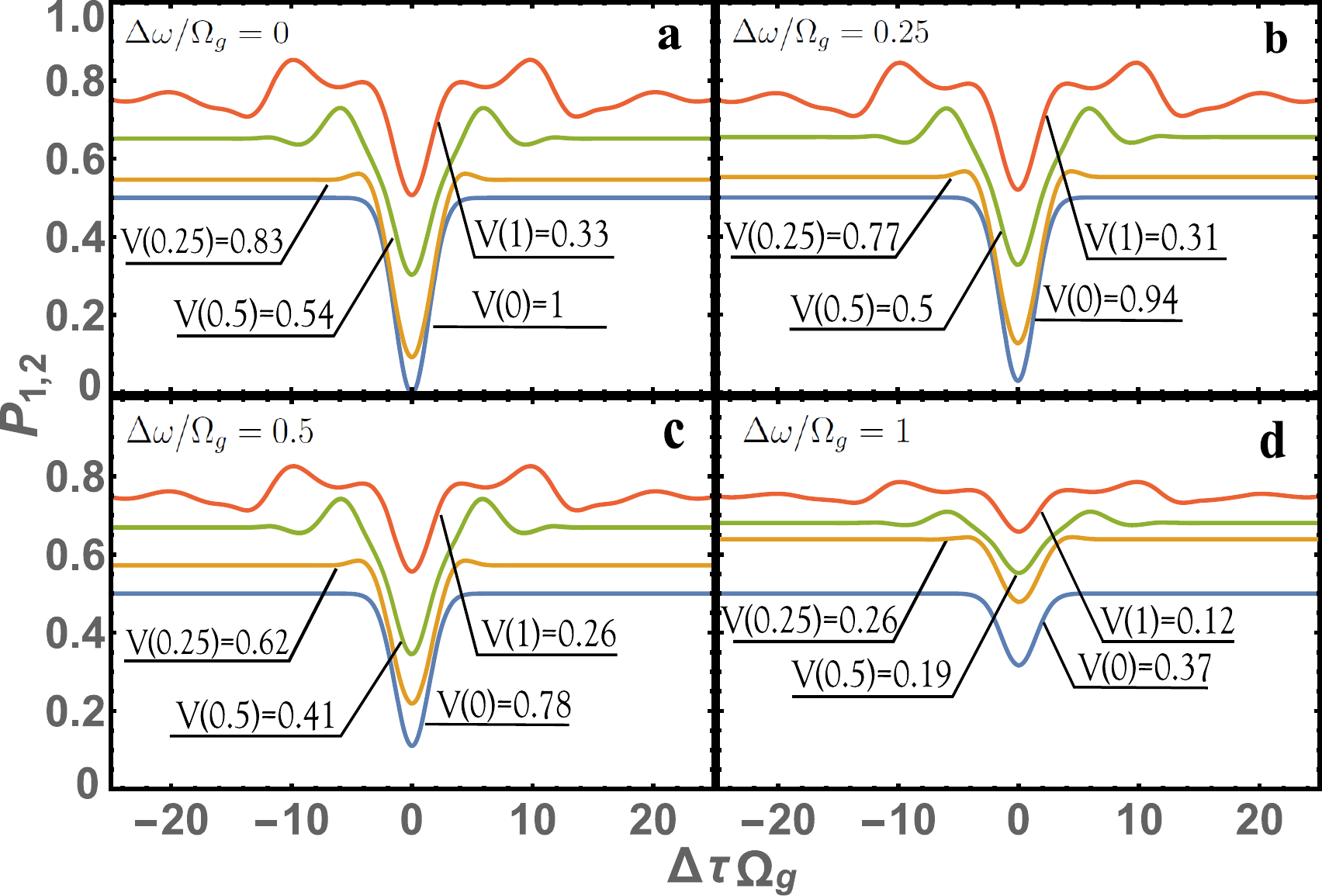}} 
\caption[fig_13]{Dependence of $ P_{1,2} $ on $ \Delta \tau \Omega_g $ (HOM dip). Case (a) corresponds to fully identical photons and cases (b), (c), (d) correspond to non-identical photons. Visibility $ {\rm V} = \frac{P_{1,2}(\Delta \tau \gg \tau_c)-P_{1,2}(\Delta \tau=0)}{P_{1,2}(\Delta \tau \gg \tau_c)}={\rm V} (\Omega_g / \Omega) $ depends on the parameter $ \Omega_g / \Omega $ (line colour corresponds to: red at $\Omega_g / \Omega=1$, green at $\Omega_g / \Omega=0.5$, brown at $\Omega_g / \Omega=0.25$, blue at $\Omega_g / \Omega=0$). The case $\Omega_g / \Omega = 0$ and visibility $ {\rm V} (0) $ corresponds to the previously known HOM interference theory with constant coefficients $ T = R = 1/2$.}
\label{fig_13}
\end{figure}

Let us present the results of calculating the value of $ P_{1,2} $ for the case
\begin{eqnarray}
\phi(\omega_1,\omega_2)=C
e^{-\frac{(\omega_1+\omega_2-\Omega_p)^2}{2\sigma^2_p}}e^{-\frac{(\omega_1-\omega_{01})^2}{2\sigma^2_1}}e^{-\frac{(\omega_2-\omega_{02})^2}{2\sigma^2_2}} .
\label{45}
\end{eqnarray}
We will be interested in the case applicable for most sources of photons $\omega_{02}-\omega_{01}\ll \omega_{01}, \omega_{02}$; $\omega_{01}/\sigma_1\gg 1$; $\omega_{02}/\sigma_2\gg 1$, in this case the normalization constant $ C=\frac{(\sigma^2_1+\sigma^2_2+\sigma^2_p)^{1/4}}{\sqrt{\pi\sigma_1 \sigma_2 \sigma_p}}$. The function (\ref{45}) allows us to analyze the value of $ P_{1,2} $ for two cases that are of practical interest. The first case is spontaneous parametric down-conversion (SPDC), for example, for $ \Omega_p = 2 \omega_0; \omega_0 = \omega_{01} = \omega_{02}; \sigma_1 = \sigma_2 = \sigma $ is SPDC of type I, where $ \sigma_p $ is the bandwidth of the pump beam, $ \omega_0 $ and $ \sigma $ are the central frequency and the bandwidth, respectively, for both the signal and the idle beams \cite{Shih_1999}. The second case/on the other hand if we consider $\sigma_{p}\rightarrow \infty$ in (\ref{45}), then this will be the case of Fock states (eg, \cite{Shih_1999,Wang_2006}). Indeed, in this case, in Eq. (\ref{45}), the $\phi(\omega_1,\omega_2)$ function will be factorized, which corresponds to Fock states. Substituting (\ref{45}) into (\ref{43}) we obtain \cite{Makarov_OL_2020, Makarov_SR_2020,Makarov_SR1_2021}
\begin{eqnarray}
P_{1,2}= \int^{\infty}_{-\infty}\biggr\lbrace e^{-(y-\frac{\Delta \omega}{\Omega_g})^2}  \left(  T^2(y)+R^2(y)\right) -
2 B e^{-(\frac{\Delta \omega}{\Omega_g})^2}T(B y)R(B y) e^{-y^2}\cos \left(B \Delta \tau \Omega_g y \right) \biggr\rbrace \frac{d y}{\sqrt{\pi}},
\label{46}
\end{eqnarray}
where $B=A\sqrt{\frac{1+\frac{\sigma^2_p}{\sigma^2_1+\sigma^2_2}}{A^2+\frac{\sigma^2_p}{\sigma^2_1+\sigma^2_2}}}$ and $ B \in (0,1) $; $A=\frac{2\sigma_1\sigma_2}{\sigma^2_1+\sigma^2_2};~\Omega_g=\sqrt{\frac{4\sigma^2_1\sigma^2_2+(\sigma^2_1+\sigma^2_2)\sigma^2_p}{\sigma^2_1+\sigma^2_2+\sigma^2_p}}$,
\begin{eqnarray}
\Delta \omega=\omega_{02}\frac{\sigma^2_p+2\sigma^2_1}{\sigma^2_1+\sigma^2_2+\sigma^2_p}-\omega_{01}\frac{\sigma^2_p+2\sigma^2_2}{\sigma^2_1+\sigma^2_2+\sigma^2_p}+\Omega_p \frac{\sigma^2_2-\sigma^2_1}{\sigma^2_1+\sigma^2_2+\sigma^2_p},
\nonumber
\end{eqnarray}
where $ T (y) $ and $ R (y) $ are determined by the Eq.(\ref{28}), with the only difference being that
\begin{eqnarray}
\Omega=\frac{4\pi n}{\omega_0 } {\bf u}_1 {\bf u}_2,~ \epsilon=\frac{\Omega_g}{\Omega}y,~\omega_0=\frac{2\sigma_p\sigma^2_1\sigma^2_2+\sigma^2_p(\sigma^2_2\omega_{01}+\sigma^2_1\omega_{02})}{4\sigma^2_1\sigma^2_2+(\sigma^2_1+\sigma^2_2)\sigma^2_p}.
\label{47}
\end{eqnarray}
If we assume that $\Omega_g / \Omega \ll 1 $, then $ T $ and $ R $ become constant quantities and such a BS becomes ``conventional'', i.e. one can always choose $ T = R = 1/2 $.  From estimates of $\Omega$ it can be seen that $\Omega_g$ can be of the order of $\Omega$, so it is necessary to take into account fluctuations of the coefficients R and T. Equation for $ P_{1,2}$, in our case Eq. (\ref{46}) at constants $ T = R = 1/2 $ is easily integrated and coincides with the known $ P_{1,2} = 1/2 ( 1-B e^{-(\Delta \omega / \Omega_g)^2} e^{- 1/4 (B \Omega_g \Delta \tau) ^ 2}) $, eg \cite{Wang_2006}. Next, consider what the value of $ P_{1, 2} $ as a function of $\Delta \tau \Omega_g $  in the case $\sigma_1 = \sigma_2 =\sigma$ for different values of $\Omega_g / \Omega $ and $\Delta \omega / \Omega_g $, but for $\Omega t_{BS} $ such that $\overline{T} = \overline{R} = 1/2 $, see. Fig. \ref{fig_13}: as $ \Omega_g / \Omega $ increases, the value of $ P_{1,2} $ tends to one.
In Fig. \ref{fig_13} also shows that when accounting for $ T $ and $ R $ from frequency, $ P_ {1,2} $ can differ significantly from the previously known HOM interference theory. This means that even if the photons are identical and the beam splitter is perfectly balanced, then $P_{1,2}$ can differ significantly from zero, which cannot be the case with ``conventional'' BS. This clarification is very important, because using a frequency-dependent BS, even in the case of identical photons and balanced BS, it is impossible to determine the degree of photon identity using the HOM interferometer.

\begin{figure}[h!]
\center{\includegraphics[angle=0,width=0.9\textwidth, keepaspectratio]{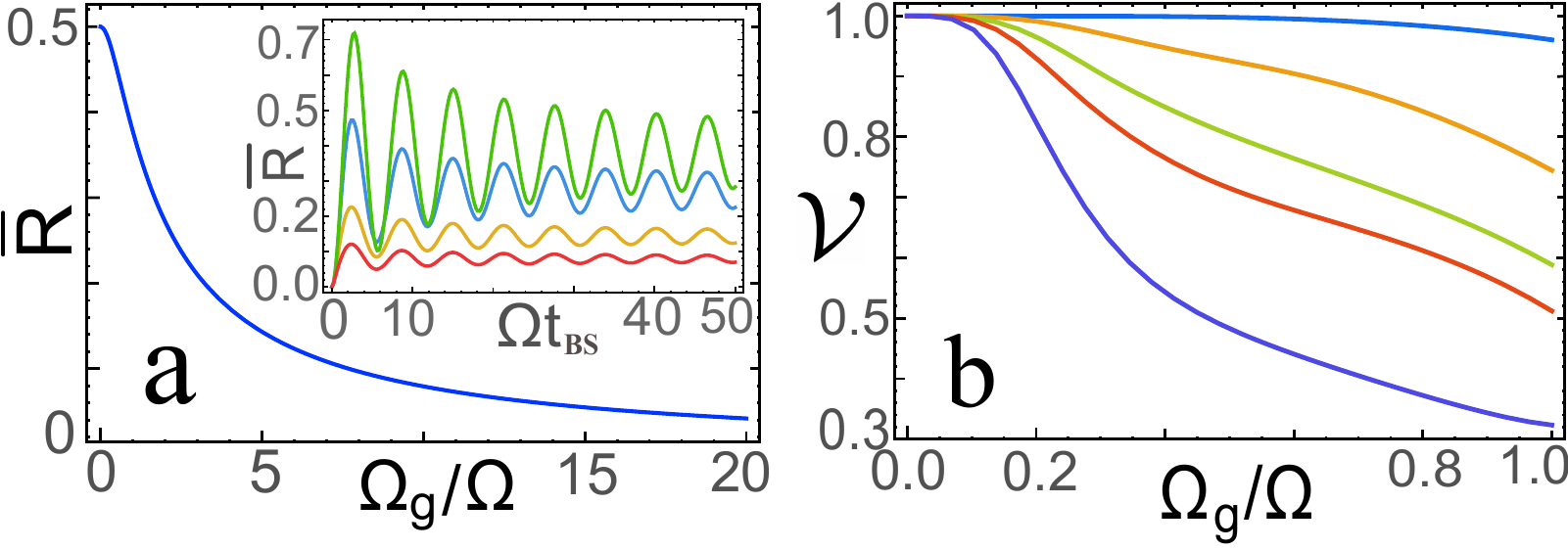}}
\caption[fig_14]{In Fig. (a) shows the dependence of the average reflection coefficient $ \overline {R} $ (see Eq. (\ref{44})) as a function of $ \Omega_g / \Omega $ for large (asymptotic) values of $ \Omega t_{BS} \gg 1 $. The internal tab contains $ \overline{R} $ depending on $ \Omega t_{BS} $ for $ \Omega_g / \Omega = 1;  2; 5 ;10$ (top-down in figure). In Fig. (b) shows the visibility of $ \mathcal{V} $ at $ \overline{R} = \overline{T} = 1/2$ depending on $ \Omega_g / \Omega $ for five values of the lengths of the coupled waveguide, which corresponds to $ \Omega t_{BS} \approx 2; 8 ; 14; 20;  30$ (top-down in figure).}
\label{fig_14}
\end{figure}

Let us give in more detail the case of the same photons and balanced BS, i.e. $ \sigma_1 = \sigma_2 = \sigma , \Delta \omega = 0 $ and $\Delta \tau =0$ the results obtained here can essentially differ from the standard HOM theory, where, as is well known, the visibility of $ V = 1 $, see Fig. \ref{fig_14}.
As can be seen from the figures, the standard HOM theory is applicable only for $ \Omega_g / \Omega \ll 1 $. Differences from standard HOM theory can be very large. This is especially noticeable in Fig. \ref{fig_14} (b), where the visibility of $V$ changes very much. You can also see Fig. \ref{fig_14}(a) (tab), that only for certain $ \Omega t_{BS} $ for given $ \Omega_g / \Omega $ you can choose $ \overline{T} = \overline {R} = 1/2 $. Moreover, the number of choices $ \Omega t_{BS} $ is limited and depends on $ \Omega_g / \Omega $, and therefore depends on the length of the coupled waveguide (length limited from above). There are no such limitations in the standard HOM theory \cite{Bromberg_2009}. For example, with $ \Omega_g / \Omega > 2 $, the HOM effect cannot be realized i.e. $ \overline{R}<1/2 $.

\section{Conclusion}
Thus, in this review paper, we considered the BS in quantum optics, as well as quantum entanglement and photon statistics at the BS output ports and the HOM effect. The special purpose of this work was to present the theory of the beam splitter in quantum optics and to systematize such BS into 2 types - these are ``conventional'' and frequency dependent BS (waveguide BS). Based on this systematization, the quantum entanglement of photons and their statistics, as well as the HOM effect, are presented. It is shown that these two types of beam splitters, despite their common matrix BS (see Eq. (\ref{3})~), can have significantly different quantum entanglement, statistics, and HOM effect. Also presented here is the theory of a frequency-dependent beam splitter based on a waveguide BS, where the reflection coefficients $R$ and the phase shift $\phi$ are found. It is shown that the special dependence of $R$ and $\phi$ (see Eq.(\ref{28})~) on the frequencies of incoming photons determines such unique properties of the waveguide BS, which differ from ``conventional'' BS. Taking into account the fact that the waveguide BS is currently acquiring an important role in quantum technologies due to the possibility of its miniaturization, this review will be useful not only for theoreticians, but also for experimenters.

\newpage
\section*{Acknowledgements}
\emph{The study was supported by the Russian Science Foundation  No. 20-72-10151; Grant of the President of the Russian Federation No. MD-4260.2021.1.2; state assignment of the Russian Federation No. 0793-2020-0005 and No. FSRU-2021-0008}

\begin{spacing}{0.7}

\end{spacing}

\end{document}